\documentclass[a4paper,11pt,preprint,superscriptaddress,nofootinbib]{revtex4-1}
\pdfoutput=1

\usepackage[T1]{fontenc}

\usepackage{empheq}
\usepackage{amsmath}
\usepackage{amssymb}
\usepackage{dsfont}
\usepackage{graphicx}
\usepackage{hyperref}
\usepackage{stackrel}
\usepackage{cancel}

\newcommand{\cA}{\mathcal{A}}
\newcommand{\cC}{\mathcal{C}}
\newcommand{\cF}{\mathcal{F}}
\newcommand{\cG}{\mathcal{G}}
\newcommand{\cH}{\mathcal{H}}
\newcommand{\cK}{\mathcal{K}}

\newcommand{\p}	{\partial}

\newcommand{\ads}{\text{AdS}}
\newcommand{\cft}{\text{CFT}}
\newcommand{\lcft}{l_\text{CFT}}

\DeclareMathOperator{\arcosh}{arcosh}
\DeclareMathOperator{\tr}{tr}

\newcommand{\bra}[1]	{\langle{#1}\vert}
\newcommand{\ket}[1]	{\vert{#1}\rangle}

\newcommand{\be}	{\begin{equation}}
\newcommand{\ee}	{\end{equation}}

\begin{document}

\title{Topological complexity in AdS$_3/$CFT$_2$}

\author{Raimond Abt}
\affiliation{Institut f{\"u}r Theoretische Physik und Astrophysik,\\ Julius-Maximilians-Universit{\"a}t W{\"u}rzburg, Am Hubland, 97074 W\"urzburg, Germany}
\author{Johanna Erdmenger}
\affiliation{Institut f{\"u}r Theoretische Physik und Astrophysik,\\ Julius-Maximilians-Universit{\"a}t W{\"u}rzburg, Am Hubland, 97074 W\"urzburg, Germany}
\author{Haye Hinrichsen}
\affiliation{Institut f{\"u}r Theoretische Physik und Astrophysik,\\ Julius-Maximilians-Universit{\"a}t W{\"u}rzburg, Am Hubland, 97074 W\"urzburg, Germany}
\author{\\Charles M. Melby--Thompson}
\affiliation{Institut f{\"u}r Theoretische Physik und Astrophysik,\\ Julius-Maximilians-Universit{\"a}t W{\"u}rzburg, Am Hubland, 97074 W\"urzburg, Germany}
\author{Ren\'e Meyer}
\affiliation{Institut f{\"u}r Theoretische Physik und Astrophysik,\\ Julius-Maximilians-Universit{\"a}t W{\"u}rzburg, Am Hubland, 97074 W\"urzburg, Germany}
\author{Christian Northe}
\affiliation{Institut f{\"u}r Theoretische Physik und Astrophysik,\\ Julius-Maximilians-Universit{\"a}t W{\"u}rzburg, Am Hubland, 97074 W\"urzburg, Germany}
\author{Ignacio A. Reyes}
\affiliation{Institut f{\"u}r Theoretische Physik und Astrophysik,\\ Julius-Maximilians-Universit{\"a}t W{\"u}rzburg, Am Hubland, 97074 W\"urzburg, Germany}
\affiliation{Instituto de F\'isica, \\Pontificia Universidad Cat\'olica de Chile, \\
Casilla 306, Santiago, Chile}

\begin{abstract}
We consider subregion complexity within the AdS$_3$/CFT$_2$ correspondence. We rewrite the volume proposal, according to which the complexity of a reduced density matrix is given by the spacetime volume contained inside the associated Ryu-Takayanagi (RT) surface, in terms of an integral over the curvature. 
Using the Gauss-Bonnet theorem we evaluate this quantity for general entangling regions and temperature.
In particular, we find that the discontinuity that occurs under a change in the RT surface is given by a fixed topological contribution, independent of the temperature or details of the entangling region. 
We offer a definition and interpretation of subregion complexity in the context of tensor networks, and show numerically that it reproduces the qualitative features of the holographic computation in the case of a random tensor network using its relation to the Ising model. 
Finally, we give a prescription for computing subregion complexity directly in CFT using the kinematic space formalism, and use it to reproduce some of our explicit gravity results obtained at zero temperature. 
We thus obtain a concrete matching of results for subregion complexity between the gravity and tensor network approaches, as well as a CFT prescription.
\end{abstract}


\maketitle

\section{Introduction}
Since the proposal of Ryu and Takayanagi \cite{Ryu:2006bv,Rangamani:2016dms} that entanglement entropy in a holographic conformal field theory (CFT) is measured by the area of minimal surfaces in asymptotically $\ads$ spacetimes, the connection between the AdS/CFT correspondence \cite{Maldacena:1997re} and quantum information has seen many exciting developments. 
In recent years, these ideas have found applications ranging from tensor networks \cite{Swingle:2009bg} and quantum error correcting codes \cite{Almheiri:2014lwa,Pastawski:2015qua} to the emergence of spacetime \cite{VanRaamsdonk:2010pw}. 

One research topic that is receiving increasing attention is the notion of complexity \cite{Papadimitrou}. 
Roughly speaking, the complexity of a pure quantum state is the minimal number of gates of any quantum circuit built from a fixed set of gates that produces this state from a given reference state. 
Complexity was first studied within the framework of the $\ads/\cft$ correspondence  in the context of time-dependent thermal state complexity, which was proposed to be dual either to the volume of the Einstein-Rosen bridge~\cite{Susskind:2014rva}, or the action of a Wheeler-DeWitt patch~\cite{BrownSusskind}. Recently, additional insight has been gained into both proposals from more detailed holographic studies~\cite{Carmi:2016wjl,Carmi:2017jqz,Kim:2017qrq}. 

\begin{figure}[b]
\begin{center}
\includegraphics[scale=0.2]{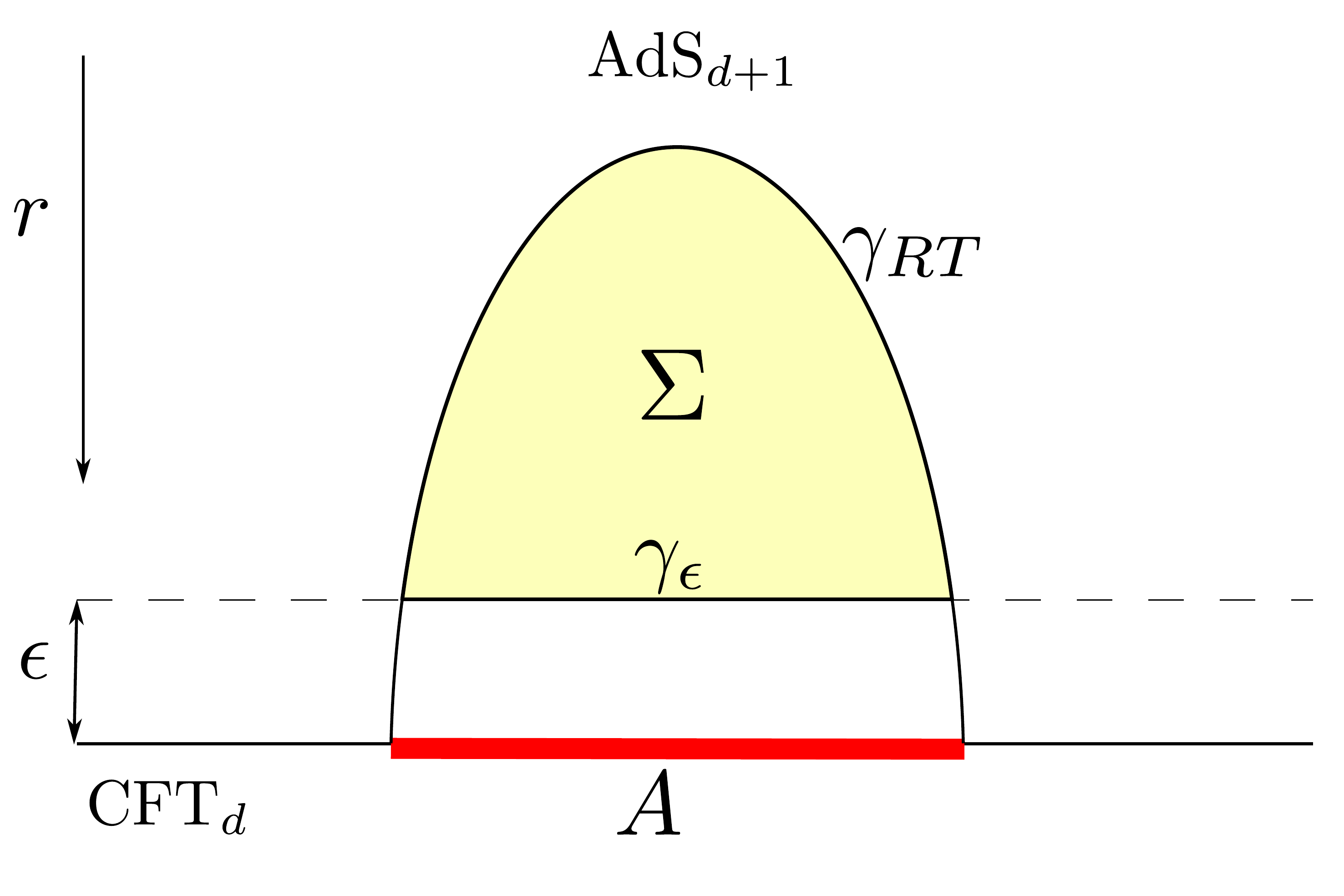} 
\vspace{-6mm}
\end{center}
\caption{The subregion complexity is computed from the regularized volume contained in the region $\Sigma$, enclosed by $\gamma_{RT}$ and the segment $\gamma_\epsilon$ of the cutoff surface.}
\label{fig:surface}
\end{figure}

The present paper is concerned with the \textit{subregion complexity} of the reduced density matrix of a finite subregion $A$. While the area of the Ryu-Takayanagi (RT) surface $\gamma_{RT}$ of $A$ is known to give the entanglement entropy of $A$ \cite{Ryu:2006bv}, it was proposed in \cite{Alishahiha:2015rta,Bakhshaei:2017qud,Roy:2017uar} that the subregion complexity should correspond to the volume of the co-dimension 1 region $\Sigma$ enclosed by $\gamma_{RT}$ and the cutoff surface (fig.~\ref{fig:surface}). Other recent proposals relate bulk volumes to \textit{Fisher information} \cite{Banerjee:2017qti} and \textit{fidelity susceptibility} \cite{MIyaji:2015mia,Alishahiha:2017cuk,Gan:2017qkz,Flory:2017ftd}.

The particular object of study of this work is the behavior of subregion complexity in $\ads_3/\cft_2$. We study a slightly different quantity than~\cite{Alishahiha:2015rta}: we define the subregion complexity of $A$ to be the integral over $\Sigma$ of the scalar curvature $R$,
\begin{align}\label{C}
\mathcal{C}(A) \equiv - \frac{1}{2}\int_\Sigma \, R \, d\sigma \,.
\end{align}
The minus sign accounts for the negative curvature of asymptotically AdS spaces.

The examples studied in the present work have constant spatial curvature, so that our definition coincides with the proposal in~\cite{Alishahiha:2015rta}. However, the definition in (\ref{C}) has several advantages. On the one hand, it is particularly natural in $\ads_3$, as the resulting quantity is dimensionless without introducing an \emph{ad hoc} scale. 
On the other hand, we will see below that, due to the appearance of $\int_\Sigma R$ in the Gauss-Bonnet theorem, the quantities of primary interest to us are determined purely by topological data. For this reason we may refer to the subregion complexity as defined by \eqref{C} as topological complexity.
Finally, the idea of defining the subregion complexity as an integral over a local \textit{complexity density} proportional to the scalar curvature is conceptually interesting on its own. 

The aim of this article is to develop and compare three complementary points of view on subregion complexity in $\ads_3/\cft_2$: within gravity, using tensor networks, and its computation using $\cft$ quantities. Our approach has two main foci. The first is the study of transitions and temperature dependence of subregion complexity from both the point of view of gravity and of tensor networks.  
On the gravity side, the Gauss-Bonnet theorem yields an elegant result: when the total length of the entangling region is held fixed, the subregion complexity~(\ref{C}) varies only by discrete jumps determined purely by the topology of $\Sigma$. 
This holds true for any number of entangling intervals at both zero and non-zero temperature, and for variations of temperature as well as the shape of the entangling region. 
In the latter case the change in complexity during topological transitions of the Ryu-Takayanagi surface is in particular independent of temperature.
While the two-interval subregion complexity was originally computed in~\cite{Ben-Ami:2016qex}, the computation for arbitrary numbers of intervals at both zero and non-zero temperature is new.

In the context of tensor network/$\ads$ proposals, we interpret the subregion complexity as the complexity of the map that optimally compresses the reduced density matrix of $A$. Using the map between random tensor networks and the Ising model proposed in \cite{Hayden:2016cfa}, by numerical simulations we reproduce the qualitative behavior at the transitions observed in gravity. 
Our numerics reproduce to a good approximation the temperature independence of the subregion complexity as measured by the volume under the Ryu-Takayanagi-surface, as well as the jump in subregion complexity when the Ryu-Takayanagi surface undergoes a topological transition for large boundary interval sizes.

Our second focus is the computation of subregion complexity within $\cft_2$. 
In continuum CFT we cannot compute complexity from first principles, because a satisfactory definition of complexity in QFT is not yet available.
(See however \cite{Hashimoto:2017fga,Chapman:2017rqy,Jefferson:2017sdb,Yang:2017nfn} for recent work in this direction.) 
Here we seek to approach this problem from a different angle by outlining the definition of a quantity using the kinematic space formalism of~\cite{Czech} which, in the case of $\cft_2$ with a holographic dual, we expect to reproduce the holographic subregion complexity for states sufficiently close to the vacuum.
We apply our prescription to compute the complexity of the vacuum when the entangling region is the entire spatial boundary, and find that it matches the gravitational computation. 
As the kinematic space measure is built from entanglement entropy, this suggests that complexity can be recovered from entanglement entropy, at least for states sufficiently close to the vacuum.
A more general computation will appear in upcoming work~\cite{Abt:2018ywl}.

The paper is organized as follows. 
In section \ref{sec:I} we consider the subregion complexity of an arbitrary number of entangling regions for locally AdS$_3$ solutions. We obtain explicit expressions from the proposal \eqref{C} for these geometries. In section \ref{sec:II} we study subregion complexity from the viewpoint of tensor networks. In section \ref{sec:III} we use the kinematic space formalism of~\cite{Czech} to define a quantity in {$\cft_2$} that, when a weakly curved gravitational dual exists, coincides with the subregion complexity. Our conclusions are summarized in section~\ref{sec:conclusions}.

\section{Subregion complexity from gravity}
\label{sec:I}
%
We begin with the computation of the subregion complexity~(\ref{C}) in asymptotically $\ads$ spacetimes of constant spatial curvature. Using the Gauss-Bonnet theorem, we derive a simple and general expression for the subregion complexity of any collection of intervals and for arbitrary geometries of constant spatial curvature. 
We illustrate this formula in detail for the specific cases of vacuum $\ads$, static BTZ black holes, and conical defect geometries. 
In vacuum $\ads$ we illustrate this formula in the case of the two-interval subregion complexity originally computed in~\cite{Ben-Ami:2016qex}, while in black holes and defect geometries we focus on the mass dependence of single interval complexity.
In general, we find that the jump in complexity that occurs when the dominant Ryu-Takayanagi surface undergoes a transition comes ``quantized'' in integer multiples of a fixed value, independent of geometric parameters of the background such as interval size or black hole temperature.

We work on constant time slices of asymptotically AdS$_3$ solutions. One first fixes an entangling region $A$ at the boundary, whose RT surface consists of geodesic(s) connecting its endpoints. 
As usual, one places a cutoff slice $\gamma_\epsilon$ near the boundary for regularization. As depicted in fig.~\ref{fig:surface}, this defines a compact two dimensional manifold $\Sigma$ with boundary $\partial \Sigma=\gamma_{RT}\cup \gamma_\epsilon$.

The Gauss-Bonnet theorem allows us to express the topological subregion complexity~\eqref{C} in a simple form: 
\begin{align}\label{GB}
\mathcal{C}(A)
=
-\frac{1}{2}\int_{\Sigma}R\,d\sigma
=
\int_{\partial \Sigma} k_g ds-2\pi \chi(\Sigma)\ ,
\end{align}
where $\chi$ is the Euler characteristic of $\Sigma$ and $ds$ is the line element along $\partial \Sigma$. The geodesic curvature $k_g$, {defined in \eqref{kg1} below}, measures how much the curve $\partial \Sigma$ deviates from a geodesic. If $\partial \Sigma$ is piecewise smooth, then $\int_{\partial \Sigma} k_gds$ is the sum of the integral along the smooth portions of $\partial \Sigma$, plus the sum of the corner angles at its turning points (where $k_g$ has delta function singularities). 

We now compute \eqref{GB} for entangling regions on $\ads_3$, BTZ black holes and the conical defects. The time slices of these solutions have constant curvature $R=-\frac{2}{L^2}$, where $L$ is the AdS radius.

\subsection{Zero temperature (AdS)}
\label{sec:IA}
\begin{figure}[t]
\begin{center}
\includegraphics[scale=0.3]{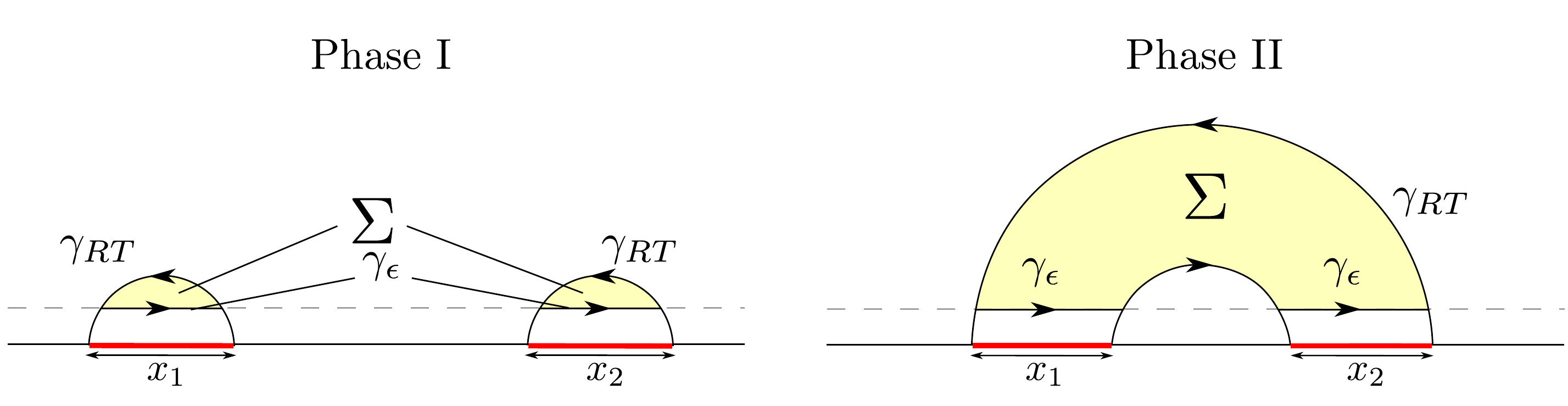} 
\end{center}
\vspace{-6mm}
\caption{The two phases of a system with two subregions. For phase I, $\Sigma$ is the union of the colored regions.}
\label{fig:phase}
\end{figure}

%
Consider first a set of two entangling intervals of lengths $x_1$ and $x_2$ in the vacuum state of a CFT$_2$, which is dual to global AdS$_3$ (fig. \ref{fig:phase}) with metric
\begin{align}\label{ds2}
ds^2=-f(r)dt^2+\frac{dr^2}{f(r)}+r^2d\phi^2\ ,
\end{align}
with $f(r)=1+\left( \frac{r}{L} \right)^2$ and $\phi\sim\phi+2\pi$.
We choose $\lcft/r$ as our defining function, corresponding to a $\cft$ metric $ds_\text{CFT}^2=\lcft^2(-L^{-2}\,dt^2+d\phi^2)$, and hence a CFT spatial circle of length $2\pi\,\lcft$.

The entanglement entropy of the two subregions is known to exhibit a transition between two configurations depending on a conformal ratio of their sizes and separation \cite{Hartman:2013mia,Faulkner:2013yia}. 
In standard analogy with statistical mechanics we refer to such competing configurations as ``phases''.
In particular, while the entanglement entropy is continuous across a transition, its first derivative jumps (this is, however, smoothed out at finite $c$).
On the CFT side, this transition can be explained as the exchange of dominance between the $s$ and $t$ channels in the four point function of the twist fields.
On the gravity side, this corresponds to the two different ways of connecting the interval endpoints by geodesics (see fig.~\ref{fig:phase}). 
In these two configurations, referred to as phase I and phase II, the total length of the corresponding geodesics is generally different. 
The RT prescription states that the actual entanglement entropy is given by the configuration for which this length is minimal, meaning that the transition occurs at the point where lengths in both phases coincide. Interestingly, {we find} that the volume of $\Sigma$ exhibits a discontinuity at the transition.

We now apply eq. \eqref{GB} to compute the {subregion complexity} in phase I. Here $\Sigma$ is the union of two disjoint regions. Since the Euler characteristic is additive, we obtain $\chi(\Sigma)=2$, since each region is topologically like a disk. 

Next we compute the integral of the geodesic curvature around the smooth parts of $\partial \Sigma$. Since the geodesics $\gamma_{RT}$ do not contribute, we only have to integrate along $\gamma_\epsilon$, which is a segment of a circle at radius $r=L\lcft/\epsilon\equiv r_\epsilon$ with $\epsilon\ll \lcft$. For a constant time slice of a metric of the form \eqref{ds2}, it is easy to show that the geodesic curvature along a circle of radius $r$ is simply 
\begin{align}\label{kg1}
k_g &= \biggl| \frac{Du}{ds} \biggr| = \frac{\sqrt{|f(r)|}}{r} \,,
\end{align}
where $u$ is the unit vector tangent to the curve. For asymptotically AdS spaces, where $f(r)\to\left( \frac{r}{L} \right)^2$ as $r\rightarrow \infty$, we obtain
\begin{align}\label{kds}
\int_{\gamma_\epsilon} k_g ds=\frac{\sqrt{|f(r_\epsilon)|}}{r_\epsilon}\int_{\gamma_\epsilon} ds=\frac{x_1+x_2}{\epsilon}+\mathcal{O}(\epsilon) \,,
\end{align}
where $x_1$ and $x_2$ denote the lengths of the intervals.
Finally, the contributions coming from the corner angles between $\gamma_{RT}$ and $\gamma_\epsilon$ have to be taken into account. Since $\gamma_{RT}$ is known to terminate perpendicularly at the boundary \cite{Rangamani:2016dms}, any joint of $\gamma_{RT}$ with $\gamma_\epsilon$ contributes with a term of $\pi/2$ to \eqref{GB} when $\epsilon\rightarrow 0$. 
Summarizing all contributions, the subregion complexity for two disjoint intervals of length $x_1$ and $x_2$ is simply given by
\begin{align}\label{}
\mathcal{C}_I(\{x_1,x_2\})=\frac{x_1+x_2}{\epsilon}-2\pi\ .
\end{align}
Similarly we can compute the {subregion complexity} in phase II, the only difference being that the Euler characteristic is now $\chi(\Sigma)=1$:
\begin{align}\label{}
\mathcal{C}_{II}(\{x_1,x_2\})=\frac{x_1+x_2}{\epsilon}\,,
\end{align}
Since both phases differ by a constant topological term
\begin{align}\label{dC}
\Delta \mathcal{C}=\mathcal{C}_{II}-\mathcal{C}_I=2\pi \,,
\end{align}
the subregion complexity exhibits a discontinuous jump at the transition, although the entanglement entropy is continuous. 
This was already computed in \cite{Ben-Ami:2016qex} by direct integration of the volume form.%
\footnote{In \cite{Gan:2017qkz} hyperbolic polygons lying in spatial slices of AdS$_3$ were considered in context of holographic complexity. The 
Gauss-Bonnet theorem was consulted to confirm their computations. Our prescription \eqref{GB} reproduces their findings as a special case.}

\begin{figure}[t]
\begin{center}
\includegraphics[scale=0.3]{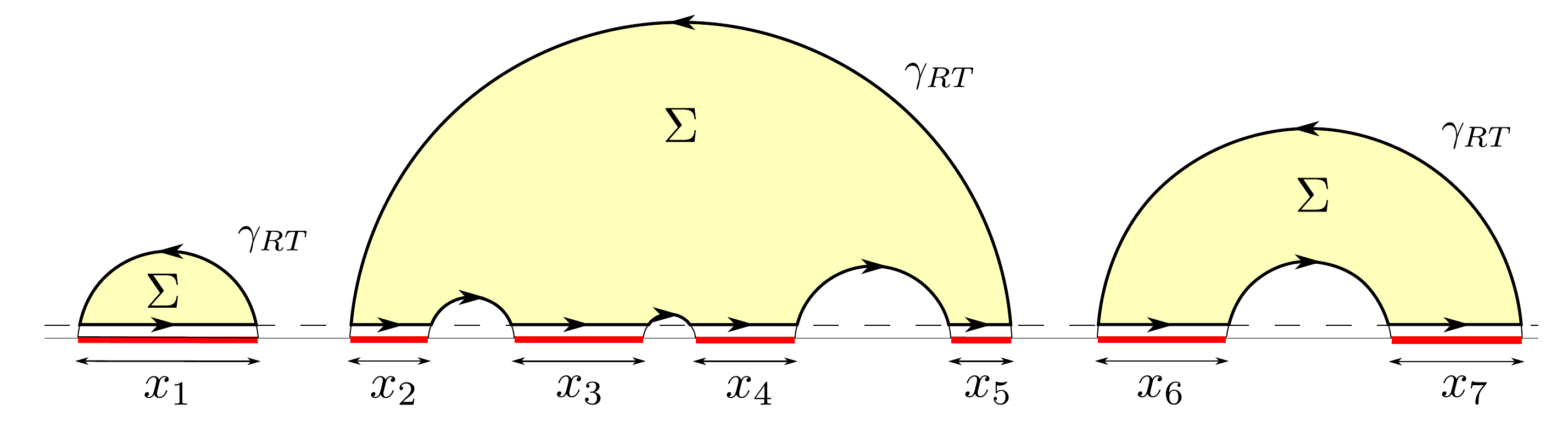} 
\end{center}
\vspace{-6mm}
\caption{Example of a configuration of RT surfaces for a several entangling intervals {($q=7$)} in the vacuum.}
\label{fig:comp}
\end{figure}

The generalization to an arbitrary number of entangling intervals is straightforward. Consider a set of $q$ disjoint intervals of length $x_i$ in the vacuum state of a {$\cft_2$} (see fig.~\ref{fig:comp}). Depending on their configuration, $\gamma_{RT}$ can take many possible forms, giving rise to various phases. 
Applying once again the Gauss-Bonnet theorem, each corner angle contributes {$\pi/2$}, hence the subregion complexity is given by 
\begin{align}\label{Cq}
\mathcal{C}(\{ x_i \})=  \frac{x}{\epsilon} +\pi q -2\pi \chi \,,
\end{align}
where $x=\sum_{i=1}^q x_i$ is the total entangling length on the boundary of the $q$ intervals, and $\chi$ the total Euler characteristic. 

At any transition of $\gamma_{RT}$, all contributions to \eqref{GB} remain the same except that the Euler characteristic $\chi$ changes, thus
\begin{align}\label{}
\Delta \mathcal{C}=-2\pi\Delta \chi\ .
\end{align}
This is the first main result of this paper: if the entangling intervals are varied while the sum of their lengths is held fixed, then subregion complexity varies discretely in multiples of $2\pi$. We shall see below that the same is true for a finite temperature state. 

In particular, the {subregion complexity} of the entire time slice of AdS$_3$ is obtained when a single entangling region covers the entire boundary circle. Setting $q=0$ (no corner angles) we obtain the result
\begin{align}\label{CAdS}
\mathcal{C}\left( \mbox{circle} \right)=2\pi \left( \frac{\lcft}{\epsilon}- 1\right) \,,
\end{align}
which will be derived in terms of CFT quantities in sec.~\ref{sec:III}.

\subsection{Finite temperature (BTZ) and conical defect}
\label{sec:IB}

We now consider a single interval of length $x$ in a CFT$_2$ on a circle at finite temperature $T$. This is dual to the BTZ black hole~\cite{Banados:1992wn}, where the metric is again of the form \eqref{ds2}, but now with
\begin{align}\label{BTZ}
f(r)=-M+\left( \frac{r}{L} \right)^2 \,,
\end{align}
where $M$ is the black hole mass (in units of $8G_N=1$) which is related to the temperature by $T=L\sqrt{M}$. It is well known that $M>0$ corresponds to black holes while $M=-1$ reproduces AdS$_3$. The geometries for $-1<M<0$ correspond to conical defects in AdS, i.e. naked singularities with no horizon. 

\begin{figure}[t]
\begin{center}
\includegraphics[scale=0.15]{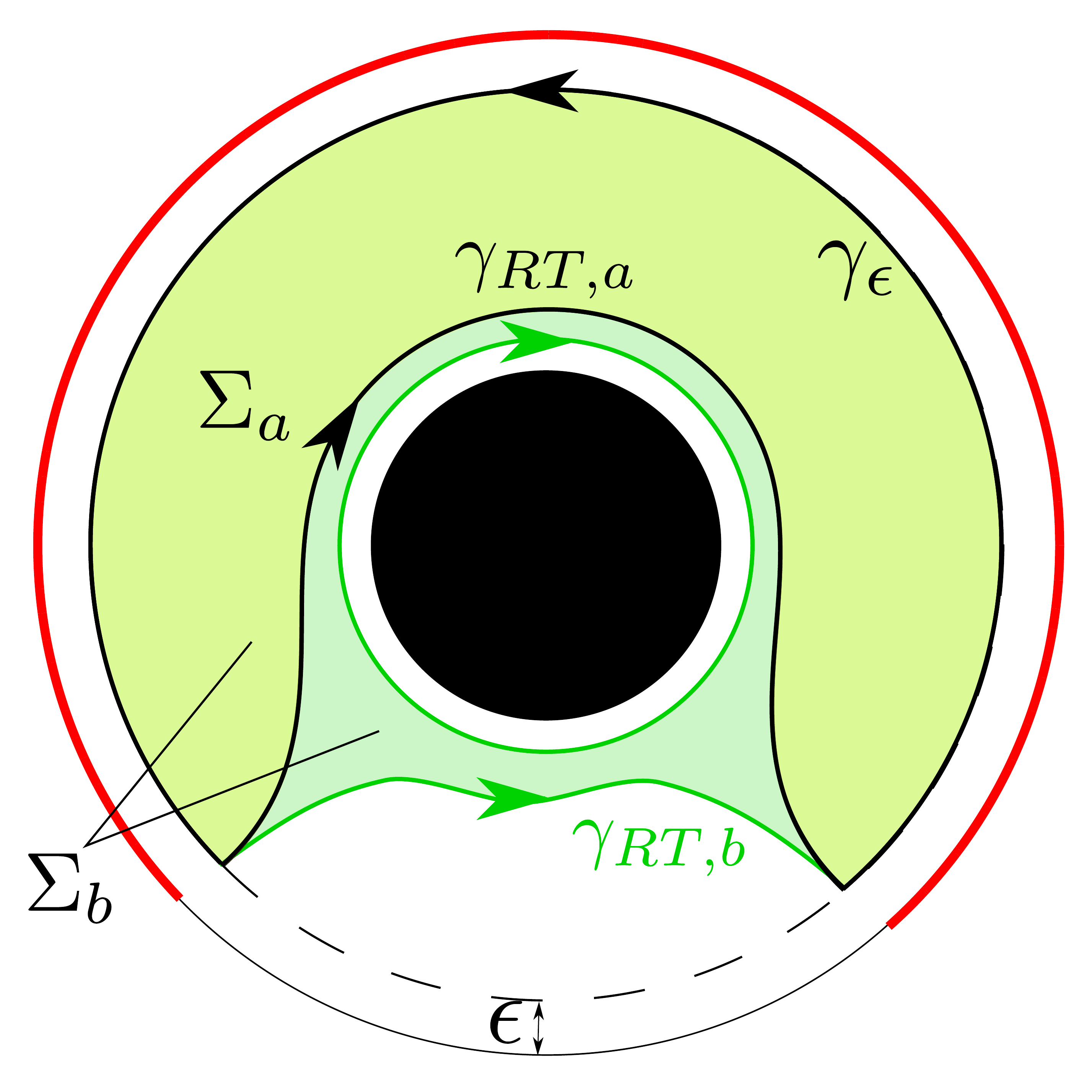} 
\end{center}
\vspace{-6mm}
\caption{The two phases $a$ and $b$ of $\gamma_{RT}$ for a single interval (red line) in the presence of a black hole.}
\label{fig:bh}
\end{figure}

In the presence of a black hole $\gamma_{RT}$ is known to exhibit two different phases $a$ and $b$, as shown in fig.~\ref{fig:bh}, provided that  the entangling region is larger than half of the boundary perimeter. In phase~$b$ the geodesic $\gamma_{RT}$ remains homotopic to the entangling region, while in phase~$a$ it is given by the geodesic of the complement plus a surface wrapping around the horizon of the black hole. Again the physically realized phase is the one where the entanglement is minimal. For low temperatures the black hole is small so that $\gamma_{RT,b}$ is shorter than  $\gamma_{RT,a}$ while for large temperatures it is the other way round. Both phases are separated by a transition point $M=M^*$ where the entanglement in both phases coincides. 

The calculation of the {subregion complexity} in each phase is analogous to the vacuum case. Indeed, in phase $a$ the result is identical to that of a single interval in the vacuum, namely
\begin{align}\label{CA}
\mathcal{C}_a(x)=\frac{x}{\epsilon}-\pi\ ,
\end{align}
independent of the mass of the black hole. 
Thus, our second main result is that the topological nature of the subregion complexity~(\ref{C}) implies it is independent of temperature. 
Although the entanglement entropy has a strong temperature dependence ($\gamma_{RT}$ changes with the black hole size), it changes in precisely such a way as to leave the volume inside constant. 
Note that this is not true in higher dimensions, and is due to the fact that the BTZ geometry is locally isometric to $\ads_3$. 

\begin{figure}[t]
\begin{center}
\includegraphics[scale=0.3]{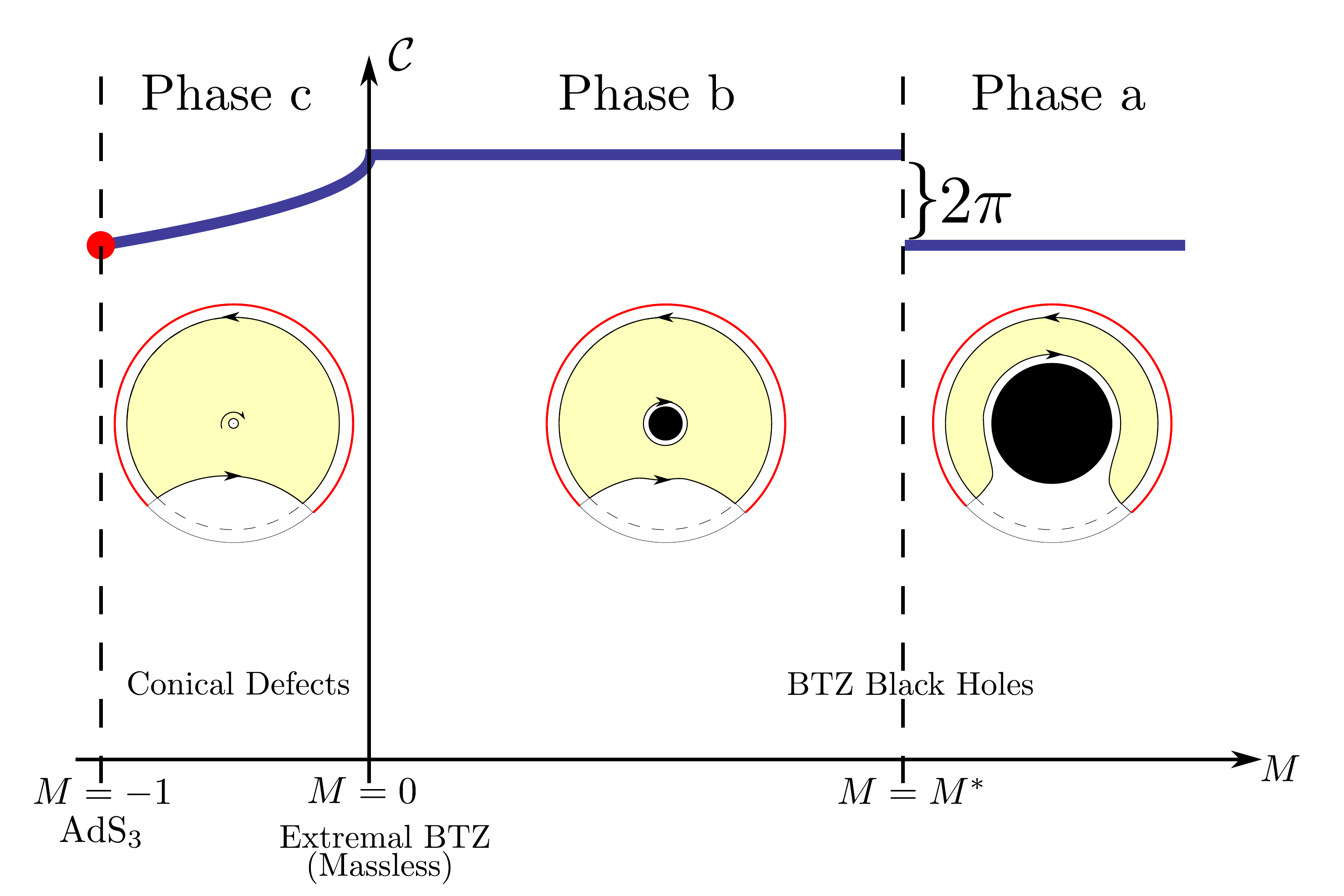} 
\end{center}
\vspace{-6mm}
\caption{Subregion complexity as function of the black hole mass, for a fixed entangling region.}
\label{bh123}
\end{figure}

As we lower the mass, the black hole gets smaller, until we hit the phase transition and pass to phase $b$, as shown in fig.~\ref{bh123}. The geodesic curvature of the horizon vanishes, and so
all contributions to \eqref{GB} remain unchanged except for the Euler characteristic, which is now $\chi(\Sigma_b)=0$, as $\Sigma_b$ is topologically an annulus. 
Therefore the corresponding complexities differ by 
\begin{align}\label{}
\Delta \mathcal{C} = \mathcal{C}_b - \mathcal{C}_a = 2\pi\,,
\end{align}
as derived earlier in \cite{Ben-Ami:2016qex} by direct integration.

As we continue reducing the temperature, we hit massless BTZ (extremal), and pass to the `naked singularity' sector. Solutions with negative $M$ correspond not to black holes, but to solutions of Einstein's equations when we place a point particle of mass $M$ at the origin. This generates a  conical defect geometry, where the deficit angle is $2\pi(1-\sqrt{-M})$. There is no horizon, and the curvature is still $R=-\frac{2}{L^2}$ everywhere except at the origin, where it has a Dirac delta peak.

The entanglement entropy for conical defects in AdS$_3$ was studied in \cite{Balasub14}. We now consider their {subregion complexity}. When the horizon disappears at $M=0$, it would seem that the topology of $\Sigma$ changes since it would no longer have any hole. However, it remains the same: given that there arises a singularity at the origin, and in order to be consistent with the homology condition for the RT surface, one must remove an infinitesimal disk around the singularity, compute the subregion complexity, and finally take the disk radius to zero.  This introduces another boundary, whose geodesic curvature is again given by \eqref{kg1} but now with $f(r)=-M+\left( \frac{r}{L} \right)^2$. The integral around the disk is 
\begin{align}\label{}
\oint k_g ds=2\pi\sqrt{f(r)}\ \underset{r\rightarrow 0}{\longrightarrow}\ 2\pi\sqrt{-M}\ \ ,\ M<0\ .
\end{align}
All other contribution to \eqref{GB} remain the same, so the {subregion complexity} for the naked singularity is
\begin{align}\label{Ccd}
\mathcal{C}=\frac{x}{\epsilon}+\pi-2\pi\sqrt{-M}\ \ ,\ M<0
\end{align}
For $M=-1$ the AdS vacuum is recovered where \eqref{Ccd} reduces to \eqref{CA}, and the {subregion complexity} again approaches the same value as in phase $a$. 

To summarize, for a single entangling region of a given size we find three different phases depicted in fig.~\ref{bh123}. Although the entanglement entropy varies non-trivially with temperature in all phases, the Gauss-Bonnet theorem ensures that the {subregion complexity} in phase $a$ and $b$ are constant, exhibiting a jump of $2\pi$ at the transition. This changes once we cross to the conical defect sector, in which a naked singularity appears which causes the {subregion complexity} to vary smoothly.   

Finally, the {subregion complexity} for $q$ intervals at finite temperature is again analogous to \eqref{Cq}, and the natural generalization of \eqref{Ccd} for the conical defect case.

\section{Tensor networks}
\label{sec:II}
%
The aim of this section is to suggest a physical interpretation of the holographic subregion complexity discussed in section~\ref{sec:I} by defining an analogous quantity for tensor networks, and to illustrate for a particular class of tensor networks that this quantity has the same qualitative behavior found in section~\ref{sec:I}.

The advantage of this approach is that tensor networks are equipped with a natural notion of complexity, allowing us to define subregion complexity explicitly for a certain class of tensor network states.
Motivated by this definition, we compute subregion complexity in random tensor networks via numerical simulations of the Ising model realization of the second R\'enyi entropy derived in~\cite{Hayden:2016cfa}.
Our simulations reproduce the qualitative behavior found in section~\ref{sec:IB} for the subregion complexity in CFT at finite temperature.

\subsection{Subregion complexity for tensor network states}
\label{sec:tn complexity}
The first goal of this section is to define an analogue of the holographic subregion complexity for network states. 
This quantity roughly measures the difficulty of building the reduced density matrix.
The remainder of this section will investigate the properties of this definition in the particular case of random tensor networks.

We begin by briefly reviewing the definition of complexity and its relation to tensor networks. 
Complexity is an information-theoretic quantity that can be defined as follows.%
\footnote{References, and a geometric approach to this problem, can be found in~\cite{2005quant.ph..2070N}.}
Starting with a Hilbert space $\cH$ with a decomposition into local units, one chooses a set $\{U_i\}$ of quantum gates, \textit{i.e.,} unitary operators acting locally on $\cH$. 
The complexity of an arbitrary unitary operator $U$ is the minimal number of gates required to represent $U$ by a product of $U_i$'s. 
The complexity of a state $\ket{\Psi}$ in $\cH$ --- or a density matrix, which can be understood as a state in $\cH\otimes\overline{\cH}$ --- is then the smallest complexity of all unitary maps sending a fixed reference state $\ket{\Psi_0}$ to $\ket{\Psi}$.

A useful definition of complexity in continuum field theory is currently unavailable. A sharper picture is, however, provided by tensor network states~\cite{EvenblyVidal:2011}. 
It is known that some tensor networks are relevant to CFT: the Multi-scale Entanglement Renormalization Ansatz (MERA) is a tensor network known to accurately approximate CFT ground states~\cite{PhysRevLett.115.180405}. 
These networks live on discretizations of hyperbolic space, and, as a result, many statements from holography have a natural realization in such states. In particular, their entanglement entropies are bounded from above by the RT formula~\cite{Swingle:2009bg}. 
Other networks can satisfy tighter bounds, for example the quantum error-correcting codes of~\cite{Pastawski:2015qua}, where it was shown that, for a single entangling interval, the discretized Ryu-Takayanagi formula holds exactly.%
\footnote{This paper applied the Gauss-Bonnet theorem to hyperbolic tesselations to make a distinct but related computation, whose aim was to quantify multipartite entanglement associated to a partition of the whole system into several components.}

As observed in~\cite{Stanford:2014jda}, the `complexity=volume' conjecture is naturally realized in tensor network constructions by associating a fixed spatial volume to each tensor. 
In this picture, one drops the focus on the fixed Hilbert space $\cH$, working instead with maps between two Hilbert spaces that are built out of tensors. 
We still require the tensors (gates) to act locally, but the output dimension is now allowed to be smaller than the input dimension.
The subnetwork of tensors $\Sigma$ connecting the (discretized) RT surface to the boundary entangling region $A$ can be interpreted as defining a map $\imath_A$ from $\cH_{RT}$, the Hilbert space of the legs cut by the RT surface, to $\cH_A$, the Hilbert space of region $A$.  
The RT surface is characterized by the property that it has the \emph{smallest} Hilbert space for any cut through the tensor network bounded by $\p A$, and the number of tensors $\cC(\imath_A)$ measures the complexity --- with respect to tensors of a given size and locality --- of the corresponding map.
In general, the resulting tensor network may not be the map with the smallest number of tensors, so that this number only constitutes an upper bound on complexity.

We define subregion complexity $\cC_A$ to be $\cC(\imath_A)$.
For certain networks --- for example, for $A$ a single interval on the boundary of a perfect tensor network~\cite{Pastawski:2015qua} --- the reduced density matrix $\rho_A$ can be recovered directly from $\imath_A$, so that $\cC_A$ describes the complexity of $\rho_A$ itself.
In general, we expect that at large central charge the complexity of building $\rho_A$ is well parametrized by $\cC_A$.

Our main interest in this section is the behavior of $\cC_A$ under transitions in the (discretized) RT surface.
Some tensor networks, such as the perfect tensor networks of~\cite{Pastawski:2015qua}, are guaranteed to exhibit discretized versions of the jumps in complexity observed in section~\ref{sec:I}. 
A more interesting illustration is given by random tensor networks~\cite{Hayden:2016cfa}, in which the RT formula is satisfied only in the limit of infinite bond dimension. 
Using numerical simulations, we will see in what follows that the qualitative behavior of the transitions is preserved at finite bond dimension.

\subsection{Random tensor networks}
%
Observables of the random tensor networks of~\cite{Hayden:2016cfa} are defined by averaging over tensor network states built from random tensors living on a fixed graph. The main result of their work is that, in a tensor network lattice with a boundary, the average value of the second R\'enyi entropy of a subregion $A$, $\overline{\mbox{tr}(\rho_A^2)}$, can be expressed as the partition function $Z_A$ of an Ising model whose boundary conditions are determined by $A$. The temperature of the Ising model vanishes as the bond dimension $D$ becomes large, in which case the Ryu-Takayanagi surface manifests as a domain wall in the spin system. In this picture, the subregion complexity of $A$ is mapped to the magnetization of the Ising model.

Let us briefly review their construction. The network lives on a graph $\Gamma$, with boundary consisting of dangling edges $\p\Gamma$. Each edge $e$ of a vertex $x$ has associated to it a vector space $\cH_e$. We assume all $\cH_e$ to have fixed dimension $D$. The tensor at vertex $x$ is a unit vector $\ket{V_x}\in\cH_x=\bigotimes_{e\in \p x}\cH_e$, whose probability distribution is invariant under unitary transformations of $\cH_x$. An edge $\langle xy\rangle$ attaching $x$ to $y$ corresponds to projection onto a maximally entangled state $\ket{xy}$ in $\cH_{xy}\otimes\cH_{yx}$. The result is a state
\be
\ket{\Psi}=\biggl( \bigotimes_{\langle xy\rangle}\bra{xy} \biggr) \cdot
\biggl( \bigotimes_{x\in\Gamma} \ket{V_x} \biggr) 
\ee
in the boundary Hilbert space $\cH_\p=\bigotimes_{e\in\p\Gamma}\cH_e$,
to which we can associate a density matrix $\rho=\ket{\Psi}\bra{\Psi}$.

Given $A\subset\p\Gamma$, we can use the swap trick to write the second R\'enyi entropy of $A$ as 
\be
e^{-S_2(A)} = \frac{\tr\bigl[ (\rho\otimes\rho) \cdot \cF_A \bigr]}
{\tr\bigl[ \rho\otimes\rho \bigr]} 
= \frac{Z_1}{Z_0} \,,
\ee
where the trace is over $\cH_\p\otimes\cH_\p$, and $\cF_A$ is the swap operator reversing the order of the tensor product in the subspace $\cH_A$.

The average value of $\rho\otimes\rho$ is found by integrating over $\ket{V_x}$.
It can be evaluated by noting that it is linear in $\ket{V_x}\bra{V_x}\otimes\ket{V_x}\bra{V_x}$, hence all we require is the average
\be
\overline{\ket{V_x}\bra{V_x}\otimes\ket{V_x}\bra{V_x}} = 
\frac{I_x + \cF_x}{D(D+1)} 
\ee
where $I_x$ and $\cF_x$ are the identity and flip operators, respectively, on $\cH_x\otimes\cH_x$.
Expand this into a sum over terms involving either $I_x$ or $\cF_x$, and define a spin variable $s_x$ which is $1$ ($-1$) if the term contains $I_x$ ($\cF_x$).
We further introduce a boundary function $h_x$ equal to $-1$ for $x\in A$, and $1$ otherwise.
For large bond dimension we can approximate 
\be
\overline{S_2(A)} = -\log\frac{\overline{Z_1}}{\overline{Z_0}} \,,
\ee
where $\overline{Z_{0,1}}$ are now expressed as partition functions
\be
\overline{Z_1} = \sum_{\{s_x\}} e^{-\cA[\{s_x\},\{h_x\}]}\,,
\quad
\overline{Z_0} = \sum_{\{s_x\}} e^{-\cA[\{s_x\},\{h_x=1\}]}\,.
\ee
The explicit form of the statistical Hamiltonian $\cA$ was derived in~\cite{Hayden:2016cfa}, and (up to constant shift) takes the form 
\be
\cA[\{s_x\}]=-\frac{1}{2}\log D\biggl(
\sum_{\langle xy\rangle} s_x s_y + \sum_{x\in\p\Gamma}s_x h_x
\biggr) \,;
\ee
this is an Ising model on $\Gamma$, whose boundary spins are held fixed to the values $\{h_x\}$.
As is well known, the most probable configuration consists of the domain wall separating two regions of opposite spins that has smallest possible length.
This domain wall coincides with the RT surface.

\subsection{Ising model reproducing the subregion complexity for BTZ black holes}
%
The analogy of Ryu-Takayanagi surfaces and domain walls of an Ising model suggested in~\cite{Hayden:2016cfa} assumes a constant bond dimension $D$, meaning that all coupling constants $J_{x;y}$ in the corresponding Ising Hamiltonian $H=-\frac12 \sum_{<x,y>} J_{x;y} s_x s_y$ take the same value $J_{x;y}=J$ and that $\beta_{\mbox{\tiny Ising}}\to \infty$ in the limit of a large central charge. In this setup the bulk geometry is taken into account by arranging the tensors (Ising spins) in such a way that their geodesic distance is constant. However, such regular tessellations are only known for few special cases as, for example, the hyperbolic plane. In particular, we are not aware of a regular equidistant tessellation of the bulk geometry in presence of a BTZ black hole.

To circumvent this problem, we suggest here that the bulk geometry can also be taken into account for any arrangement of the Ising spins by assigning non-constant couplings in such a way that the energetic cost of domain walls reproduces the correct geodesic length. In the following we consider the example of a non-spinning BTZ black hole, reproducing the transition between the phases $a$ and $b$ illustrated in fig.~\ref{fig:bh}. The generalization to other examples is straight forward.

To reproduce the transition in the BTZ case, we first map the standard coordinates $(r,\phi)$ in a constant-time slice of the metric (\ref{ds2}),(\ref{BTZ}) to conformal coordinates $(\eta,\phi)$ with $\cos\eta=T/r$ defined on a rectangle $\eta\in[0,\pi/2),\pi\in[-\pi,\pi)$, turning the metric into
\begin{equation}
ds^2=\sec^2(\eta)\, \bigl( L^2 \,d\eta^2+T^2 \,d\phi^2\bigr).
\end{equation}
In these coordinates, the black hole horizon and the conformal boundary correspond to $\eta=0$ and $\eta=\pi/2$. 

\begin{figure}[t]
\begin{center}
\includegraphics[width=85mm]{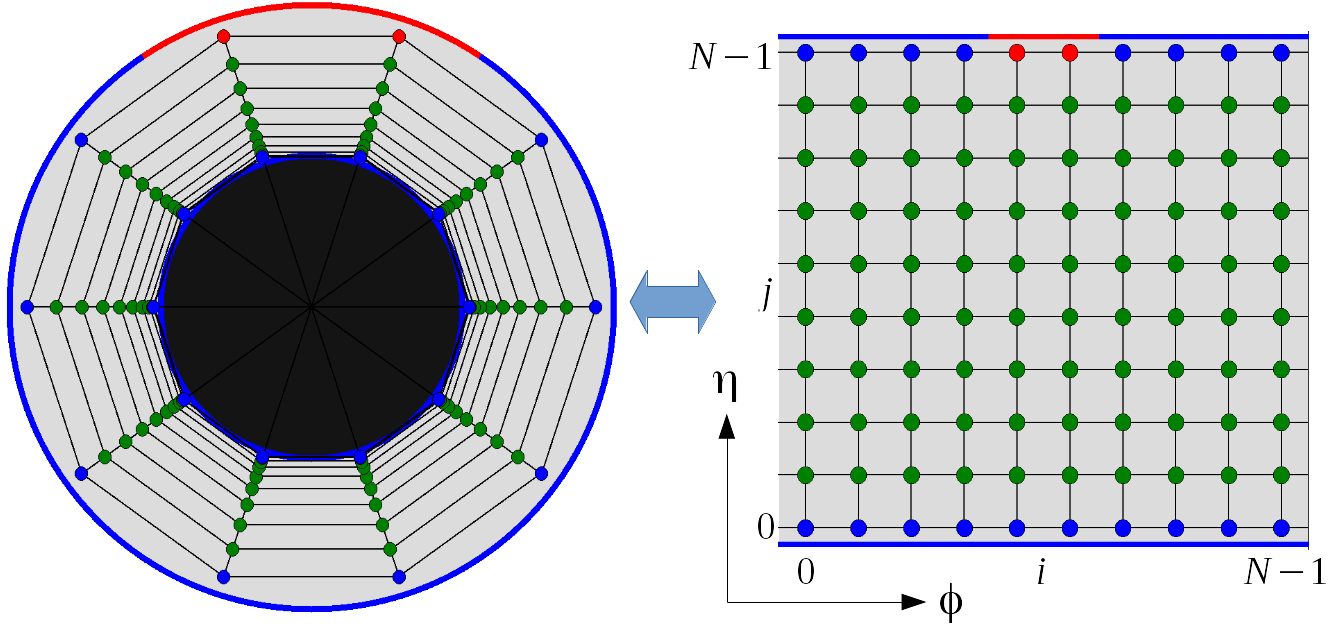}
\caption{\label{figIsingMap} BTZ black hole (left) with radial coordinate $\arctan(r/L)$ mapped to conformal coordinates $\phi,\eta$ (right) where an Ising model on a square lattice is embedded. The top and the bottom row of spins are fixed according to the respective boundary conditions (red=$\uparrow$, blue=$\downarrow$) while the green spins are allowed to fluctuate.}
\end{center}
\end{figure}

Next, we embed a square lattice of $N \times N$ Ising spins in this rectangle, as sketched in fig.~\ref{figIsingMap}. Labeling the lattice sites by two indices $i,j\in\{0,1,\ldots,N-1\}$, the spins $s_x = s_{i,j}=\pm1$ are located at
\begin{equation}
\phi = 2\pi \frac{2i+1}{2N}\,,\qquad
\eta = \frac\pi 2\frac{2j+1}{2N}\,.
\end{equation}
At the top and the bottom row of the lattice the Ising spins are fixed by the boundary conditions
\begin{equation}
s_{i,j} := \begin{cases} -1 & \text{if } j=0 \\
          \text{sign}\Bigl(\frac{N \phi}{4 \pi} -\bigl|i-\frac{N}{2}\bigr|\Bigr)  & \text{if } j=N-1\,, 
         \end{cases}
\end{equation}
where $\phi=x/L$ is the size of the entangling region.

Each horizontal (angular) bond cuts a vertical line element of a domain wall with $\Delta\phi=0,\Delta\eta=\pi/2N$, corresponding to the geodesic length $\Delta s \approx L \sec(\eta) \Delta \eta$. Likewise each vertical (radial) bond cuts a horizontal line element $\Delta\phi=2\pi/N,\Delta\eta=0$ with the geodesic length $\Delta s \approx T \sec(\eta) \Delta\phi$, where $\eta$ corresponds to center of the bond. Thus, assigning the coupling constants
\begin{equation}
\begin{split}
\text{horizontal: }
J_{i,j;i+1,j} \;:=\; \frac{\pi L}{2N} \sec\Bigl(\frac\pi 2\frac{2j+1}{2N}\Bigr) \\
\text{vertical: }
J_{i,j;i,j+1} \;:=\; \frac{2 \pi T}{N} \sec\Bigl(\frac\pi 2\frac{j+1}{N}\Bigr)\,,
\end{split}
\end{equation}
the energy contribution of a domain wall is approximately proportional to its geodesic length while the total magnetization would reflect the enclosed volume. However, it should be noted that the rotational invariance of the Ising model is broken on a square lattice at low temperatures, preferring domain walls that are aligned with the lattice. As we will see below, this causes the simulated domain wall to deviate slightly from the analytically expected RT surface. Nevertheless, the results from the Ising model can be used as a good approximation which qualitatively reproduce the results derived above.

\begin{figure}[t]
\begin{center}
\includegraphics[width=85mm]{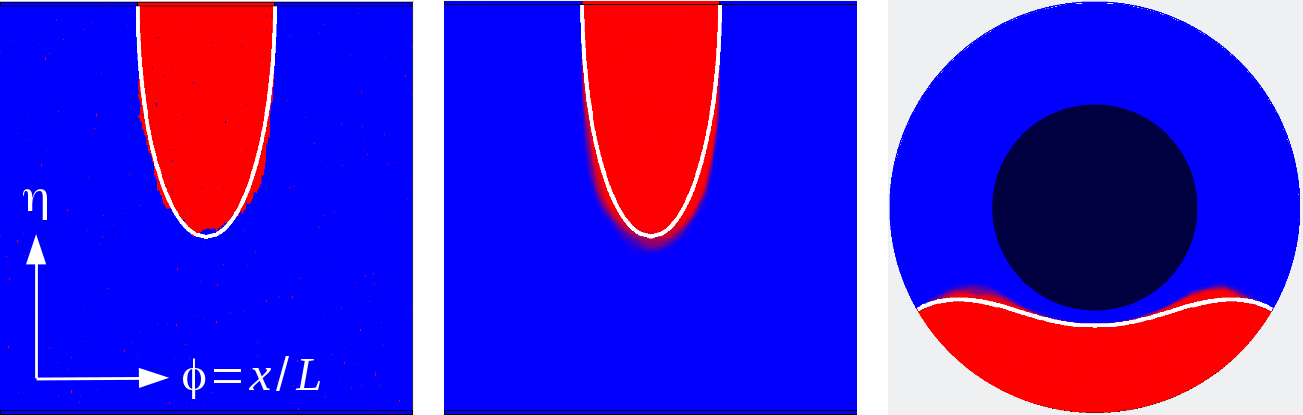}
\caption{\label{figDemoIsing} Left: Snapshot of a typical configuration of Ising spins on a lattice with 256x256 sites near equilibrium. The white curve marks the theoretically expected RT surface. Center: Magnetization averaged over many independent configurations. Right: Average magnetization mapped back to the Poincare disk.}
\end{center}
\end{figure}

\subsection{Simulation results}
\begin{figure}
\begin{center}
\includegraphics[width=85mm]{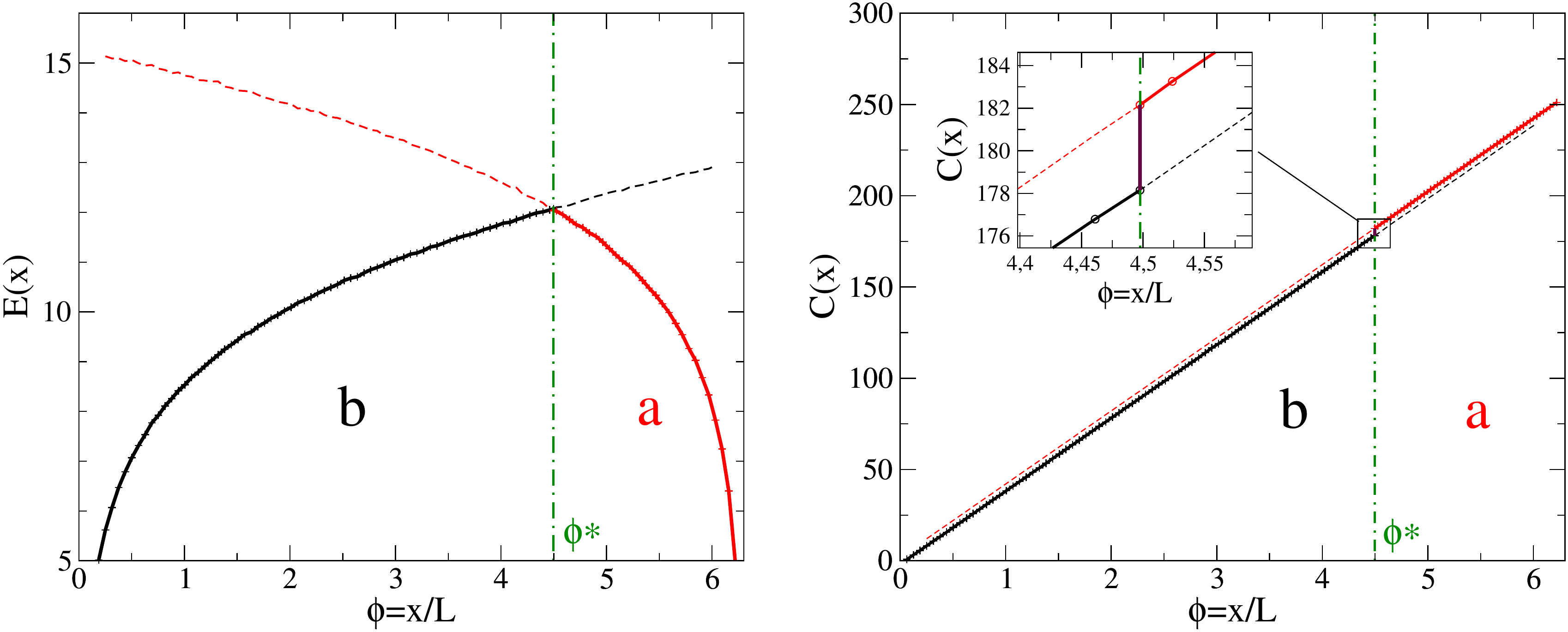}
\caption{\label{figVaryingAngle} Numerical results on a lattice with $200\times 200$ sites for a BTZ black hole with mass $M=0.1$. Left: Numerically measured entanglement of the two solutions as functions of the subregion size. As can be seen, the lines cross precisely at the theoretically expected transition point, marked by the vertical green dashed line. Right: Corresponding complexity, reproducing the linear law. The inset shows a magnification where the discontinuous jump occurs.}
\end{center}
\end{figure}

%
In order to find the equilibrium configuration of the Ising model numerically, we use a standard heat bath dynamics at very low temperature. First we measure the entanglement $E(x)$ for a constant mass $M$ and varying subregion size $\phi=x/L$. In the Ising model the entanglement is given by the energy difference
\begin{equation}
E(x) \approx -\frac12 \sum_{<x,y>} J_{x;y} (s_x s_y-1)
\end{equation}
which has to be compared with the exact result 
\begin{equation}
\begin{split}
E_b(x) &= 2\,\log \Bigl[\frac{2L}{\sqrt M \epsilon} \sinh\Big(\frac{\sqrt M x}{2 L}\Bigr)\Bigr]\\
E_a(x)&=E(2\pi L-x)+2\pi\sqrt M\,
\end{split}
\end{equation}
in the phases $a$ and $b$. Here $\epsilon$ is the cutoff distance of~$\gamma_\epsilon$ at the conformal boundary which is expected to scale with the lattice spacing. The transition takes place at a subregion size $x^*$ where $E_a(x^*)=E_b(x^*)$, giving
\begin{equation}
x^* = L \phi^* = -\frac{L}{\sqrt{M}}\log\Bigl(1-\tanh(\pi\sqrt M)\Bigr).
\end{equation}
As shown in the left panel of fig.~\ref{figVaryingAngle}, the numerically estimated entanglement in the two phases reproduces the expected behavior and the two curves intersect accurately at the expected value of $x^*$, which is marked as a vertical dashed line in the figure.

Next we compute the complexity $\cC_{a,b}(x)$ in the two phases. As can be seen in the right panel, the numerical results nicely reproduce the predicted linear law $\cC(x)=x/\epsilon+const$. Both lines are separated at a constant distance, leading to a discontinuous jump of the complexity at the transition. However, the magnitude of the jump $\Delta \cC \approx 4.0 \pm 0.3$ is clearly smaller than $2\pi$, reflecting the limitations of the model. 

\begin{figure}
\begin{center}
\includegraphics[width=85mm]{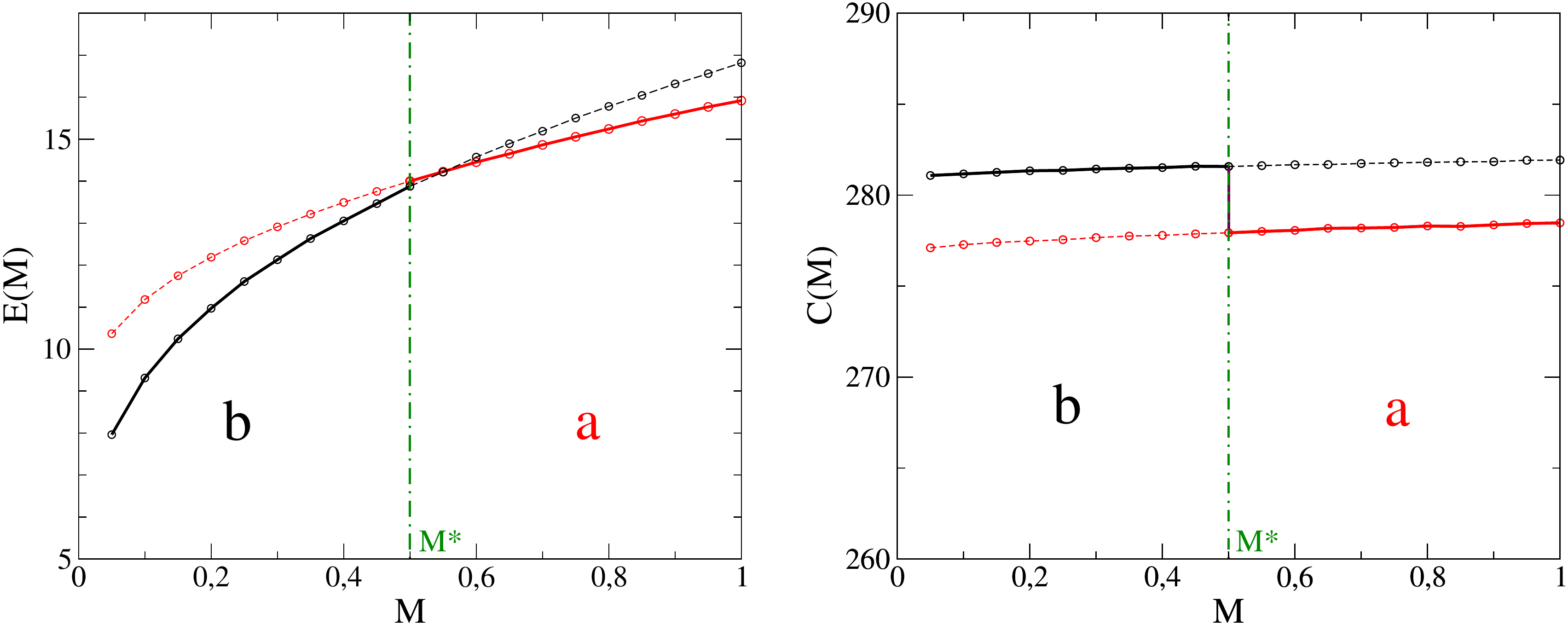}
\vspace{-2mm}
\caption{\label{figVaryingMass} 
Left: Numerically determined entanglement for varying mass measured on a lattice with $120^2$ sites and a fixed subregion size $\phi=x/L=5.31946$ for which the transition is expected to take place at $M=0.5$. Right: Numerically determined complexity as a function of the mass, where the implicit mass dependence of the cutoff $\epsilon$ has been removed (see text), reproducing Fig.~\ref{bh123} for $M>0$.
}
\end{center}
\end{figure}

Finally, we repeat the simulation for varying mass $M$ between 0 and 1 for a fixed subregion size. Here one has to take into account that the lattice implicitly determines the cutoff $\epsilon$ and that it varies with $M$. In order to determine $\epsilon$, we compare the integrated bulk volume with the total sum of site volumes on the lattice. Then we subtract the expected influence of the cutoff left and right of the expected transition point. The result is shown in fig.~\ref{figVaryingMass}. As can be seen, the 
simulations fairly reproduce the finding that the complexity is independent of the mass. At the transition the jump with $\Delta\mathcal C = 3.8(3)$ is again smaller than $2 \pi$.  

So far we cannot explain why $\Delta \mathcal{C}$ deviates from $2\pi$. The deviation could be related to the fact that the use of a square lattice at low Ising temperature breaks the rotational symmetry of the Ising model, so that domain walls aligned with the lattice and diagonal ones behave differently. This is reminiscent of the situation in the MERA, where the minimal cut line is not unique.

\section{Subregion complexity in field theory}
\label{sec:III}
%
The subregion complexity of~\cite{Alishahiha:2015rta} comprises a refinement of the holographic ``complexity equals volume'' (C$=$V) proposal~\cite{Susskind:2014rva}, in the sense that it depends on a choice of entanglement region $A$, while reducing to the C$=$V proposal when $A$ is all of space.
However, there is as yet no satisfactory independent definition of subregion complexity in field theory, leaving open the question of what, if anything, this quantity tells us about field theory.
Section~\ref{sec:II} offered a picture in the tensor network language for how this quantity should be interpreted: as the number of tensors required to compress the information contained in $\rho_A$ to a Hermitian operator acting on a Hilbert space associated to the Ryu-Takayanagi surface.

In this section, we address this problem from a different perspective.
Rather than demanding a definition from first principles, we ask instead, how does one \emph{compute} the subregion complexity within field theory?
Our proposal is this: for states sufficiently close to the $\ads$ vacuum, the subregion complexity can be computed from the entanglement entropy using the \emph{kinematic space} formalism~\cite{Czech}.
In this section we verify this proposal for the simplest case, where the state is the vacuum and the entangling interval is the entire spatial slice.%
\footnote{Recent work~\cite{Miyaji:2016mxg,Czech:2017ryf} has proposed a boundary prescription for reproducing the ``complexity equals action'' proposal within field theory in terms of the Liouville action, and \cite{Czech:2017ryf} develops a relationship to kinematic space. As of yet it is not clear how to generalize this approach to subregion complexity, nor what its explicit relation to our prescription is. This is an interesting direction for future study.}
Results for excited states and non-trivial entangling intervals will appear in upcoming work~\cite{Abt:2018ywl}.

The goal of~\cite{Czech} was to derive a CFT expression for the perimeter of an arbitrary bulk region using \emph{kinematic space} $\cK$, the space of geodesics of the constant time slice. 
When the bulk is a weakly curved dual to a large-$c$ CFT, the Ryu-Takayanagi prescription gives a correspondence between points in kinematic space and entangling regions in the CFT. Using the differential entropy of \cite{Headrick:2014eia}, \cite{Czech} showed that the perimeter can be expressed as the integral over a region in kinematic space, with respect to a measure defined in terms of the entanglement entropy. Here, we extend this result (in the case of vacuum $\ads$) by expressing the bulk volume in terms of an integral, with respect to an appropriate measure, over a region in $\cK\times\cK$.

We begin by reviewing the construction of \cite{Czech}. We then propose an expression for the volume in terms of entanglement entropy, which we apply to pure AdS$_3$.%
\footnote{We restrict our attention to pure AdS since the kinematic space has been worked out in great detail. However, there has been recent work on kinematic spaces corresponding to conical defects \cite{Cresswell:2017mbk} and black holes \cite{Zhang:2016evx,Asplund:2016koz}.}
Evaluating our expression explicitly in the case where the entangling region covers the entire boundary space, we find that it agrees with equation~(\ref{CAdS}).

\subsection{Kinematic space}
\begin{figure}[t]
\begin{center}
\includegraphics[scale=0.15]{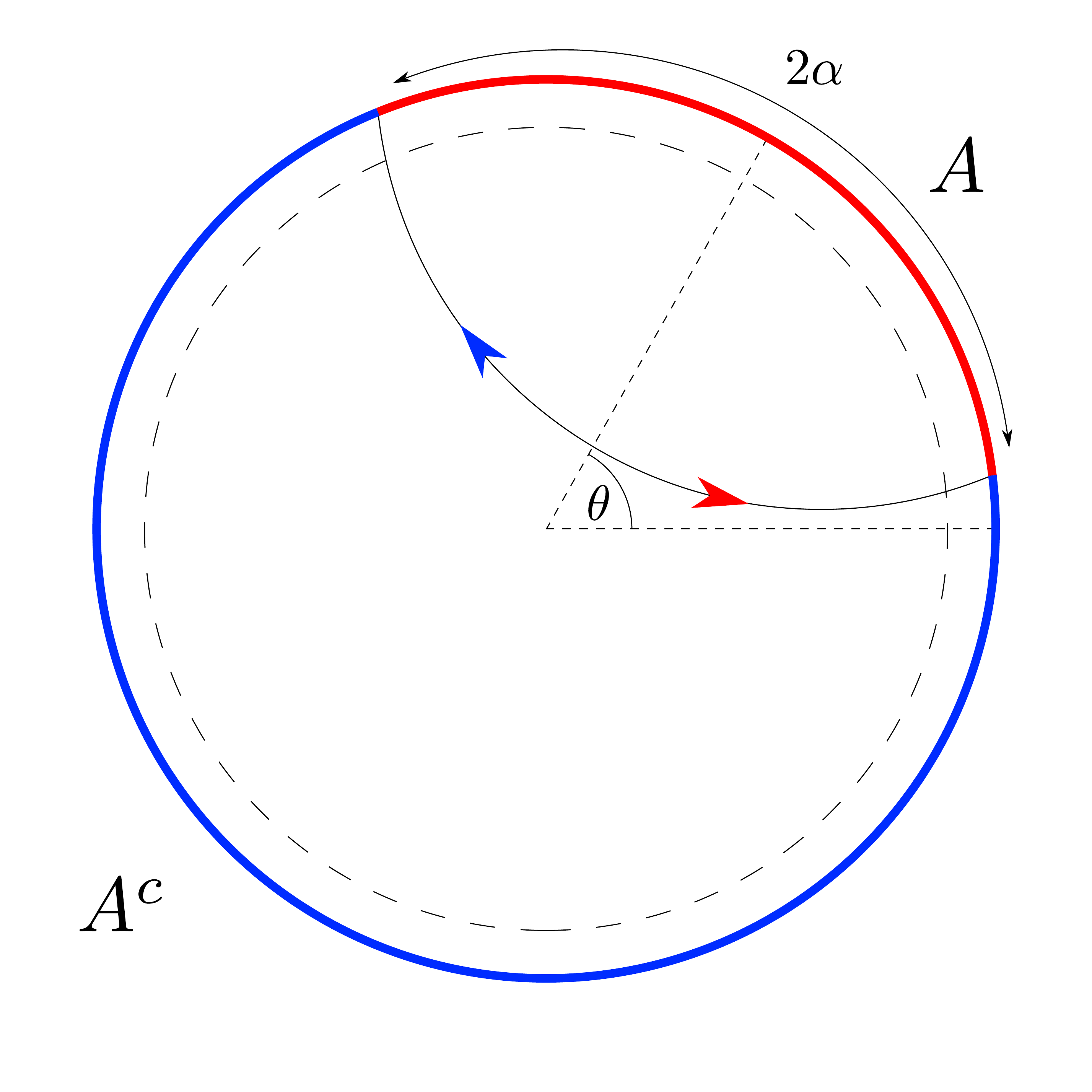} 
\vspace{-5mm}
\end{center}
\caption{Oriented geodesics. We associate the geodesic given by
			$(\theta,\alpha)$ with the entangling region $A$ and
			the orientation given by the red arrow. The geodesic
			$(\theta+\pi,\pi-\alpha)$ on the other hand is associated
			with $A^c$ and the orientation given by the blue arrow.
			We have chosen the orientation of geodesics in such a
			way that it matches the one assigned to them in section
			\ref{sec:I}.	
			}
\label{fig: oriented geod}
\end{figure}

%
We write the metric of a spatial slice of vacuum $\ads$ in the form
\begin{equation}
 ds^2=L^2(d\rho^2+{\sinh^2}\rho\, d\phi^2) \,,
\end{equation}
related to the coordinates in~\eqref{ds2} by $\sinh\rho=r/L$. The geodesics are parametrized conveniently by
\begin{equation}\label{geo}
 \cos(\alpha)=\tanh(\rho)\cos(\phi-\theta) \,,
 \quad
 \alpha\in (0,\pi) \,, \; \theta\in S^1 \,,
\end{equation}
where $2\alpha$ is the opening angle of the geodesic and $\theta$ the center of the region subtended by the geodesic (fig.~\ref{fig: oriented geod}). Each complete oriented geodesic is specified uniquely by a pair $(\theta,\alpha)$, making this a global coordinate system on the kinematic space $\cK$. Note that the orientation reversal of the geodesic $(\theta,\alpha)$ is $(\pi+\theta,\pi-\alpha)$.

An oriented geodesic in the bulk is naturally associated to an entangling interval $(u,v)$ in the CFT, where
\begin{equation}
  u = \theta-\alpha \,, 
  \qquad 
  v = \theta+\alpha \,.
\end{equation}
Flipping the orientation of a geodesic is thus associated with exchanging the entangling region with its complement (fig. \ref{fig: oriented geod}). In the limit $\alpha\rightarrow 0$ the geodesic $(\theta,\alpha)$ shrinks to a  point on the boundary of the spatial slice. We may therefore identify the boundary $(\theta,\alpha=0)$ of kinematic space with the circle on which the CFT lives. Note that we work with the same metric on the spatial circle as in section~\ref{sec:I},  $ds_{S^1}^2=\lcft^2d\phi^2$. 

\begin{figure}[t]
\begin{center}
\includegraphics[scale=0.25]{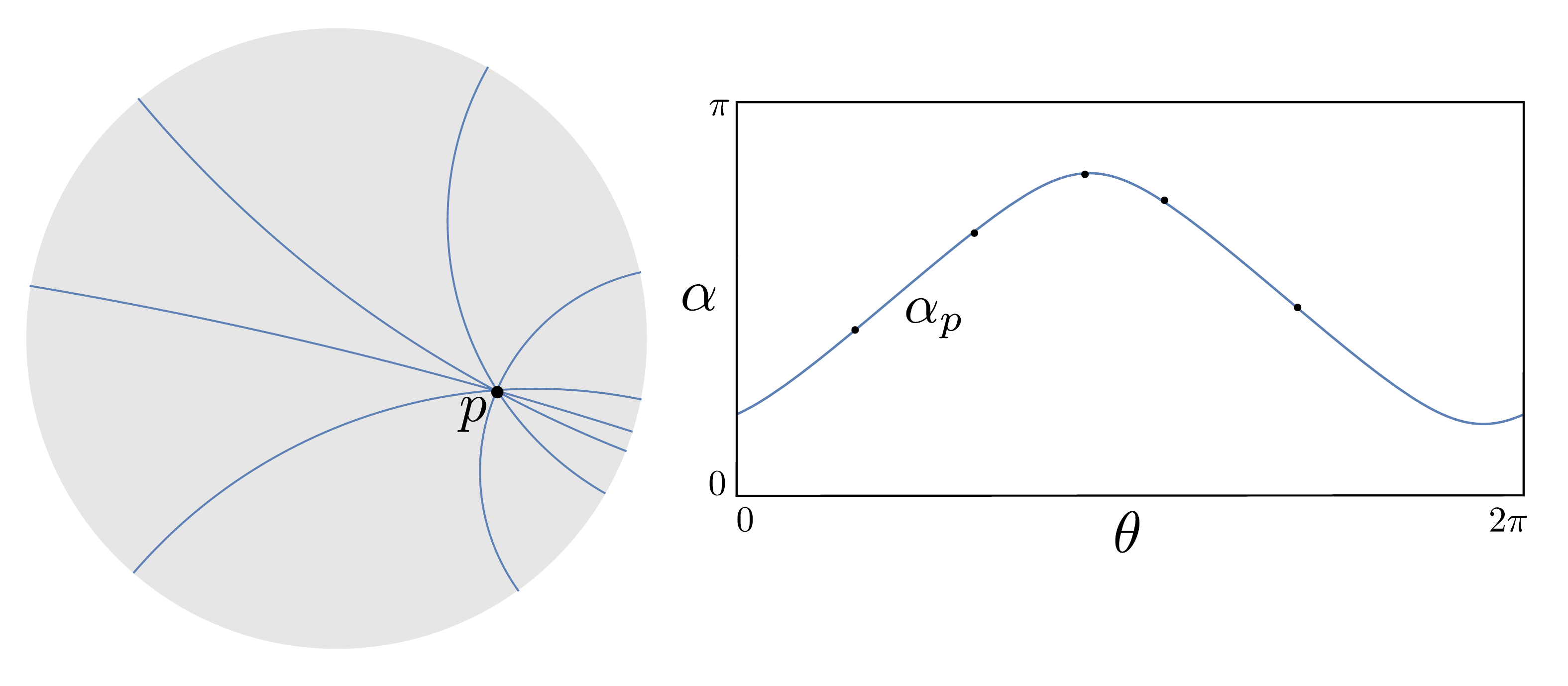} 
\end{center}
\vspace{-5mm}
\caption{Point curves in kinematic space. A given point $p$ in AdS
		(LHS) is associated with a point curve $\alpha_p(\theta)$ in kinematic
		space (RHS).
		This point curve is formed by all geodesics
		that intersect $p$.
		This graphic was generated using \cite{Template} 
		with permission of the authors.
		}
\label{fig: point curve}
\end{figure}

Our discussion will make extensive use of the concept of a \textit{point curve}, the one-parameter family of geodesics passing through a point. Given a point $(\rho,\phi)$, its point curve is the set of $(\theta,\alpha)$ in $\cK$ satisfying~\eqref{geo}. Each point in the $\ads$ spatial slice is therefore encoded by a point curve in $\cK$ (see fig.~\ref{fig: point curve}).
%

To recover information about the geometry of $\ads$ requires a geometry on $\cK$.
The required object is the \textit{density of lines}, which is a volume form on kinematic space~\cite{Czech}
\begin{equation}\label{density}
\begin{split}
	\omega
	=
	\frac{\partial^2S}{\partial u\partial v}du\wedge dv
	=
	-\frac{1}{2}\partial^2_\alpha S \, d\theta\wedge d\alpha
	=
	\frac{c}{6}\frac{1}{\sin^2(\alpha)}d\theta\wedge d\alpha
	\ .
\end{split}
\end{equation}
Here, $c$ is the central charge of the dual CFT, while $S$ is the field theory entanglement entropy on circle of length $2\pi\,\lcft$ obtained by Cardy and Calabrese \cite{Calabrese:2009qy}
\begin{align}
	S 
	&= \frac{c}{3}\log\biggr(
		 \frac{2\lcft}{\epsilon}\sin\bigl(\frac{v-u}{2}\bigr)\biggl) \nonumber\\
	&= \frac{c}{3}\log\biggr(\frac{2\lcft\sin(\alpha)}{\epsilon}\biggl) \,.
\label{EE}
\end{align}

\begin{figure}[t]
\begin{center}
\includegraphics[scale=0.25]{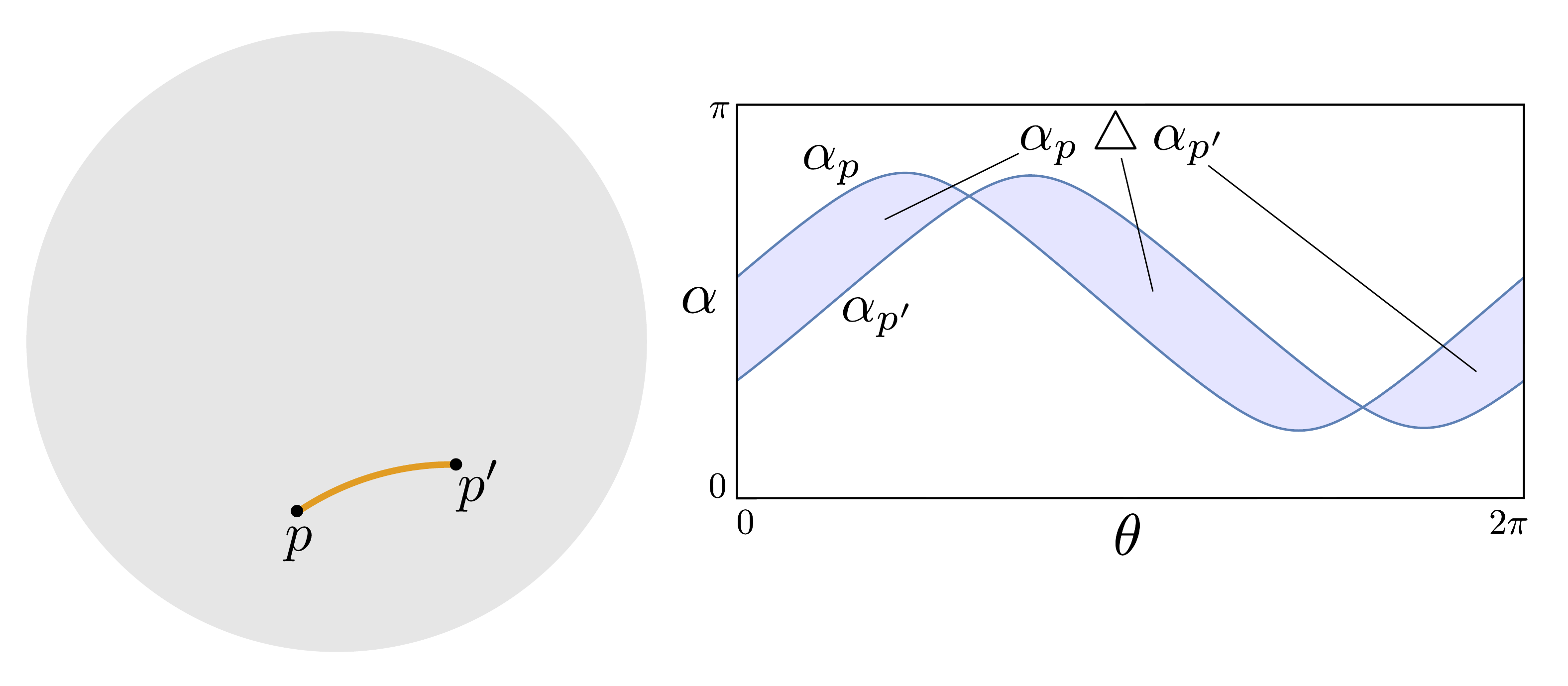} 
\end{center}
\vspace{-5mm}
\caption{The distance between two points $p,p'$ in the bulk (LHS)
			is given by an integral over the region enclosed by the
			point curves $\alpha_p,\alpha_{p'}$ (RHS).
			This graph was generated using \cite{Template}
		    with permission of the authors.
			}
\label{fig: Kinematic length}
\end{figure}

Equipped with the volume form $\omega$, kinematic space now allows us to reconstruct bulk geometric objects from CFT entanglement entropies. This was done in \cite{Czech} for the length of bulk curves: to any bulk curve $\gamma$ we can associate a two-dimensional region $\cG_\gamma$ of $\cK$ consisting of the geodesics intersecting $\gamma$. 
Using the differential entropy of~\cite{Headrick:2014eia}, it was shown that the length of $\gamma$ is proportional to the integral, with respect to the measure $\omega$, of the intersection number of the geodesic with $\gamma$. 
For instance, the geodesic distance $\lambda(p,p')$ between two points $p,p'$ of the spatial slice is given by the integral
\begin{equation}\label{length}
 \frac{\lambda(p,p')}{4G_N}=\frac{1}{4}\int_{\alpha_p\bigtriangleup\alpha_{p'}}\omega \,.
\end{equation}
The integration region $\alpha_p\bigtriangleup\alpha_{p'}$ is the region bounded by the two point curves $\alpha_p(\theta)$ and $\alpha_{p'}(\theta)$ of the points $p$ and $p'$, respectively, and is comprised of all geodesics intersecting the geodesic arc between $p$ and $p'$ (fig.~\ref{fig: Kinematic length}).

Of course, our interest in this paper is in the computation of bulk volumes. In what follows we will illustrate how to compute bulk volumes in terms of chord lengths and therefore, using \eqref{length} and \eqref{density}, in terms of entanglement entropy.

\subsection{Bulk volumes at zero temperature}
\begin{figure}[t]
\begin{center}
\includegraphics[scale=0.15]{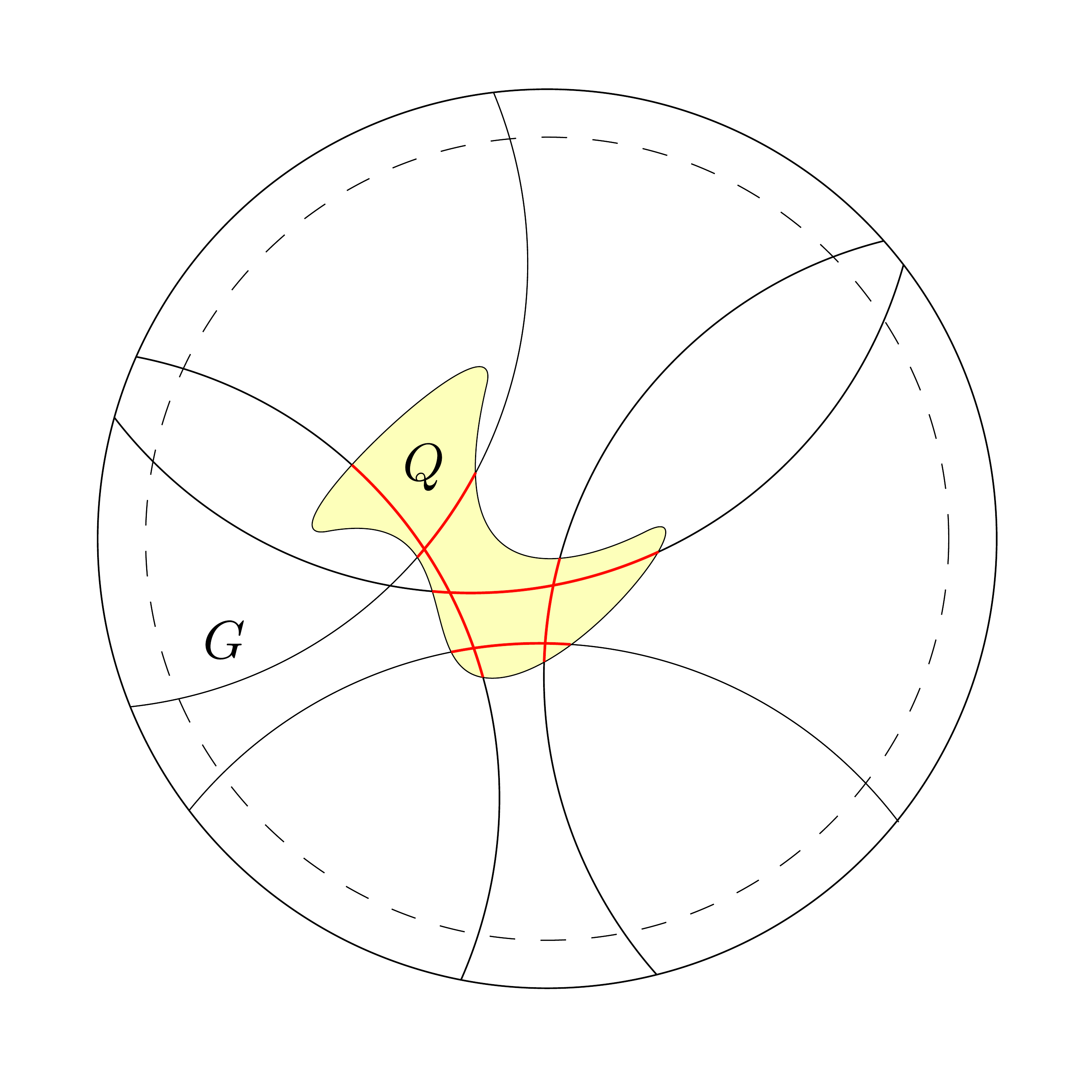} 
\end{center}
\vspace{-7mm}
\caption{In order to obtain the volume of some set $Q$ in terms of kinematic space quantities, we formulate it as an integral over all geodesics G that intersect $Q$. The chords $G\cap Q$ are red colored.}
\label{fig: geodesics and Q}
\end{figure}

%
Equation \eqref{length} is based on the Ryu-Takayanagi formula expressing the length of a geodesic in terms of the entanglement entropy of the corresponding region,
\begin{equation}
	\frac{\ell}{4G_N}=S \,,
\end{equation}
so we may also write
\begin{equation}
	\omega
	=
	-\frac{1}{8G_N}\partial^2_\alpha \ell\, d\theta\wedge d\alpha \,.
\end{equation}
We compute the volume of a bulk region $Q$ as follows. In analogy with~\eqref{length}, we wish to express it as an integral over all geodesics intersecting $Q$ (fig. \ref{fig: geodesics and Q}).
%
%
We will see that the correct expression is 
\begin{equation}
\label{eq: general Kin volume}
	\frac{vol(Q)}{4G_N}
	=
	\frac{1}{2\pi}\int_{\mathcal{G}_Q}\!\lambda_Q\,\omega \,.
\end{equation}
Here $\cG_Q$ is the set of all geodesics $G\in\cK$ intersecting $Q$, while $\lambda_Q(G)$ is the length of the intersection $G\cap Q$  (depicted in red in fig.~\ref{fig: geodesics and Q}). Observe that $\lambda_Q(G)$ is an integral over $\cK$ (see \eqref{length}). Therefore \eqref{eq: general Kin volume} is an integral over $\cK\times\cK$. 

General expressions of this type are known in the integral geometry literature (see \textit{e.g.} chapter~17 of \cite{Santalo}). Let us briefly sketch how to prove \eqref{eq: general Kin volume}.  The first step is to confirm the formula for disks centered at the origin. This can be done via an explicit calculation that we show below. The second step is to verify certain properties of volumes, such as additivity, for the integral on the right hand side of \eqref{eq: general Kin volume}.
They allow us to generalize the validity of \eqref{eq: general Kin volume} to arcs of annuli. 
In the infinitesimal limit we recover the volume element of the $(\rho,\phi)$ coordinates, allowing us to recover the Riemann integral from \eqref{eq: general Kin volume} for arbitrary regions. We will present a more detailed discussion of this proof in future work \cite{Abt:2018ywl}.

\begin{figure}[b]
\begin{center}
\includegraphics[scale=0.15]{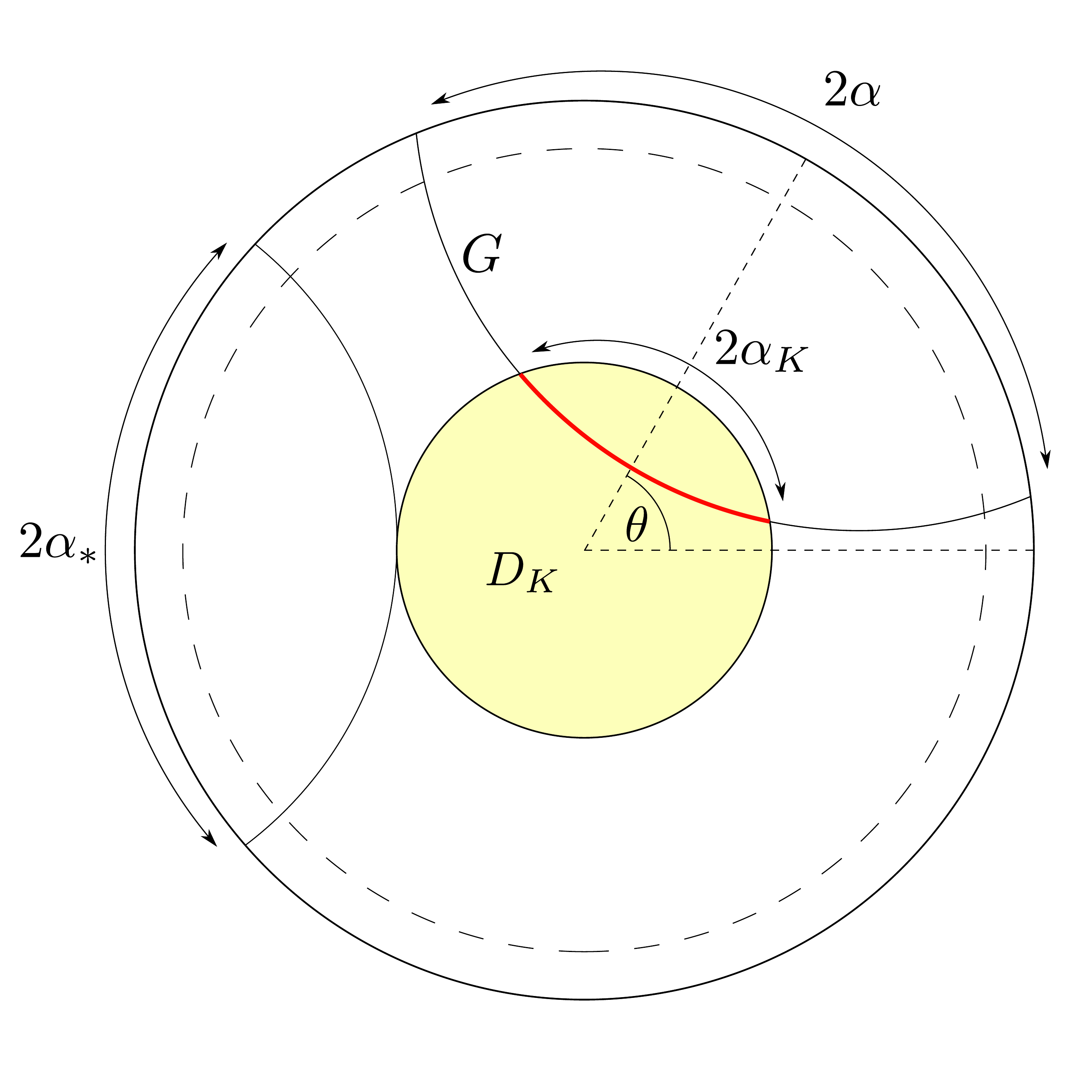} 
\end{center}
\vspace{-5mm}
\caption{To calculate the volume of the disk $D_K$ we need to consider all geodesics $G$ that intersect it. 
Their opening angle $2\alpha$ introduces another opening angle $2\alpha_K$ on the
boundary of the disk in a natural way. 
The angle $\alpha_*$ corresponds to a geodesic that is tangential to $D_K$.}
\label{fig: kin space disc}
\end{figure}

\medskip
We now show that \eqref{eq: general Kin volume} holds for
disks $D_K$ of radius $K$ around the origin (fig.~\ref{fig: kin space disc}).
%
%
The chord length is straightforward to compute and takes the form
\begin{equation}
 \lambda_{D_K}(\theta,\alpha)
 =
 L \arcosh \Bigl( 1+2\sinh^2(\rho_K)\sin^2(\alpha_K) \Bigr) \,,
\end{equation}
where $\sinh(\rho_K)=K/L$ and $2\alpha_K$ is the opening angle of the geodesic $(\theta,\alpha)$  on the boundary of $D_K$ (fig.~\ref{fig: kin space disc}). $\alpha_K$ is related to the boundary angle $\alpha$ via
\begin{equation}
\label{eq: angles bdy and cutoff}
	\tanh(\rho_K)\cos(\alpha_K)
	=
	\cos(\alpha) \,.
\end{equation}
As we sketched above, to establish~\eqref{eq: general Kin volume} it suffices to prove 
\begin{equation}
	vol(D_K)
	=
	-\frac{1}{4\pi}
	\int_0^{2\pi}d\theta\int_{\alpha_*}^{\pi-\alpha_*}d\alpha\,
		\lambda_{D_K}\partial_\alpha^2\ell \,,
\label{eq:explicit vol}
\end{equation}
where $\alpha_*$ corresponds to a geodesic tangent to the boundary of $D_K$ (fig.~\ref{fig: kin space disc}). Geodesics with a smaller opening angle do not intersect $D_K$ and therefore do not contribute. This $\alpha_*$ is given by \eqref{eq: angles bdy and cutoff} with $\alpha_K$ set to zero,
\begin{equation}
	\tanh(\rho_K)
	=
	\cos(\alpha_*) \,.
\end{equation}
It is convenient to express \eqref{eq:explicit vol} as an integral over $\alpha_K$. Using
\begin{equation}
\label{eq: koordinate trafo}
	\partial_\alpha^2\ell\ d\alpha
	=
	\partial_{\alpha_K}^2\lambda_{D_K}d\alpha_K
\end{equation}
leads to
\begin{equation}
\label{eq: calc of volume}
\begin{split}
vol(D_K)
	&= 
	-\frac{1}{4\pi}\int_0^{2\pi}d\theta\int_{\alpha_*}^{\pi-\alpha_*}
	d\alpha\, \lambda_{D_K}\partial_\alpha^2\ell
	\\
	&=
	-\frac{1}{2}\int_0^\pi d\alpha_K
		\lambda_{D_K}\partial^2_{\alpha_K}\lambda_{D_K}
	\\
	&=
	2\pi L^2\Big(\sqrt{1+\frac{K^2}{L^2}}-1\Big)
\,,
\end{split}
\end{equation}
reproducing the well-known result for the volume of the disk $D_K$ and thereby confirming \eqref{eq:explicit vol}.

\smallskip
As a special case of \eqref{eq:explicit vol}, \eqref{eq: calc of volume} can be directly compared to the complexity of the entire CFT circle~\eqref{CAdS} that we derived using the Gauss-Bonnet theorem. Since the scalar curvature $R$ is constant, the expression \eqref{C} for complexity is proportional to the volume of the time slice. This complexity is computed by \eqref{eq: calc of volume} with $K$ set to the cutoff radius $r_\epsilon=L\,\lcft/\epsilon$:
\begin{equation}
\label{eq: approx volume}
\begin{split}
	\mathcal{C}(\mbox{circle})
	&
	=
	-\frac{1}{2}R\ vol(D_{r_\epsilon})
	\\
	&
	=
	\frac{2G_N}{\pi L^2}
	\int_{\mathcal{G}_{D_{r_\epsilon}}}\lambda_{D_{r_\epsilon}}\omega
	\\
	&
	=
	2\pi\Big(\frac{\lcft}{\epsilon}-1\Big)+\mathcal{O}(\epsilon) \,.
\end{split}
\end{equation}
This successfully reproduces the complexity computed in \eqref{CAdS}.

We can now combine the volume formula \eqref{eq: general Kin volume} with the formula for distances \eqref{length} to obtain a volume formula for any region $Q$ in terms of entanglement entropy. 
To do so we apply \eqref{length} to the chord length $\lambda_Q$:
\begin{equation}
\label{eq:volume ito entropy}
\begin{split}
	\frac{vol(Q)}{4G_N^2}
	=
	&
	\frac{1}{2\pi}\int_{\mathcal{G}_Q}\Big(
		\int_{\alpha_p\bigtriangleup\alpha_{p'}}\omega
		\Big)\omega
	\\
	=
	&
	\frac{1}{8\pi}\int_{\mathcal{G}_Q}d\theta d\alpha
	\int_{\alpha_p\bigtriangleup\alpha_{p'}}d\theta'd\alpha'
		\partial_\alpha^2S(\alpha)\partial_{\alpha'}^2S(\alpha')\ .
\end{split}
\end{equation}
Here $p(\theta,\alpha)$ and $p'(\theta, \alpha)$ are the
points where the geodesic $(\theta,\alpha)$ intersects the boundary of $Q$. Note that we have assumed that each geodesic intersects the boundary of $Q$ exactly twice, so that \eqref{eq:volume ito entropy} holds only for $Q$ convex. 

Equation \eqref{eq:volume ito entropy} is the main result of this section.
It computes the volume of any convex region $Q$, and thus reproduces the subregion complexity upon setting $Q=\Sigma$ (which is always convex).
Therefore, \eqref{eq:volume ito entropy} constitutes an explicit expression for the holographic subregion complexity purely in terms of CFT quantities, namely entanglement entropies.

We emphasize that this result is not the end of the story.
While \eqref{eq:volume ito entropy} will reproduce the volume of any convex bulk region, determining a valid integration region without knowledge of the bulk region it corresponds to is in general a difficult problem. 
Fortunately, in the case of an entangling region, which is bounded by geodesics, this problem is considerably simplified. 
Explicitly verifying the general formula \eqref{Cq} is an interesting problem, whose details will be presented in upcoming work~\cite{Abt:2018ywl}.
More generally, it would be helpful to understand this problem in the case of finite temperature and for time slices of non-constant curvature.
We will address these questions in part in~\cite{Abt:2018ywl}.

\section{Conclusions and outlook}
\label{sec:conclusions}
%
In this paper we took steps toward understanding the properties of subregion complexity in CFT$_2$ from three points of view: the original holographic proposal of~\cite{Alishahiha:2015rta}, the definition and study of a tensor network analogue, and a prescription for computing it directly within $\cft_2$. 

Within gravity, we studied a modified ``topological complexity'' proposal.
Using the fact that for locally hyperbolic spaces, the curvature $R = - 2/L^2$ is constant, we rewrote the holographic volume proposal as an integral of the curvature scalar.
Such a definition of complexity density may be reflective of the loss of degrees of freedom along an RG flow. 

For the case of AdS$_3$/CFT$_2$, the new form is readily evaluated using the Gauss-Bonnet theorem, giving a simple universal formula~(\ref{Cq}) valid for an arbitrary number of entangling intervals and at any temperature. 
Particularly interesting was the change in complexity during transitions between topologically distinct RT surfaces. 
At these transitions, the subregion complexity jumps by a discrete quantity proportional to the Euler characteristic of a bulk region bounded by geodesics.
In particular, the jump comes in integral multiples of a basic unit ($2\pi$ in our normalization), irrespective of the geometry of the entangling region or black hole temperature.
Surprisingly, our result also implies that complexity in the black hole background is independent of the size of the black hole, and hence of temperature. 

Interesting questions for the future include generalizing this approach to higher dimensions; understanding subregion complexity using the optimization approach of \cite{Miyaji:2016mxg,Caputa:2017urj,Caputa:2017yrh,Czech:2017ryf}; 
relating our approach with the holographic renormalization properties of the different proposals for complexity \cite{Carmi:2016wjl,Carmi:2017ezk}; and studying subregion complexity in time-dependent systems \cite{Carmi:2017jqz}. 

Turning to tensor network states, we proposed that their subregion complexity should be understood as the number of local tensors required to build the map embedding the the Hilbert space cut by the RT surface in the Hilbert space of the entangling region $A$. 
The observed jumps in holographic subregion complexity are then understood to arise from qualitative jumps in the form of the optimal compression of $\rho_A$ to a Hilbert space of smaller dimension. 
We studied this complexity for the random tensor networks of~\cite{Hayden:2016cfa} in the presence of a black hole using its map to an Ising model.
Using numerical computations we reproduce the discontinuous jump of
subregion complexity in this approach, although our numerical value $\Delta \mathcal{C}=4.0 \pm 0.3$ differs from the gravity result $\Delta \mathcal{C}=2 \pi$. We leave it as an interesting question for future research to track down the origin of this discrepancy. Reassuringly, fig.~\ref{figVaryingMass} displays independence on temperature to a good approximation. The non-vanishing but small slope is another limitation of this model deserving further investigation.

Finally, we gave a prescription for computing the subregion complexity directly in CFT based on the kinematic space formalism. Our prescription expresses complexity as an integral built from entanglement entropies. We showed that, at zero temperature, our formula coincides with the gravity result, and verifyied the computation explicitly in the case $A=S^1$. If the subregion complexity proposal of \cite{Alishahiha:2015rta} provides a useful measure of the complexity of the reduced density matrix, our results suggest a deeper relationship between complexity and entanglement. To investive this relation further, it will be necessary to gain a deeper understanding of the field theory interpretation of subregion complexity. 
Another interesting generalization is to extend our prescription to finite temperature by working with kinematic space for black hole geometries. The case of arbitrary entanglement regions in the vacuum and at finite temperature will be presented in upcoming work~\cite{Abt:2018ywl}.

It is promising that we find coinciding results for the subregion complexity in concrete examples from three perspectives --- gravity, tensor networks, and CFT --- and we are optimistic that we will see more progress along these lines in the near future.

\begin{acknowledgments}
We wish to thank Souvik Banerjee, Stefan Fredenhagen, Pascal Fries, Daniel Harlow, Debajyoti Sarkar and Tadashi Takayanagi for fruitful discussions. 
The work of CMT is supported by a Research Fellowship from the Alexander von Humboldt Foundation. 
IR was partially funded by Conicyt PCHA \#2015149744.
\end{acknowledgments}

\bibliographystyle{JHEP}
\bibliography{complexity}

\end{document}